\documentclass[aps,twocolumn,showpacs,preprintnumbers,nofootinbib,prl,superscriptaddress,groupedaddress,10pt]{revtex4-1}

\makeatletter
\def\l@subsubsection#1#2{}
\def\l@subsubsubsection#1#2{}
\makeatother

\usepackage{graphicx,amssymb,amsmath,amsthm,amsfonts,epsfig,epsf}
\usepackage[usenames,dvipsnames,svgnames,table]{xcolor}
\usepackage{epstopdf}
\definecolor{darkred}{rgb}{0.5,0,0}
\usepackage{aas_macros}
\usepackage{bm}
\usepackage{dcolumn}
\usepackage[latin1]{inputenc}
\usepackage{latexsym}
\usepackage{rotating}
\usepackage{longtable}

\usepackage{enumerate}
\usepackage{url}
\usepackage[linktocpage]{hyperref}

\def\be{\begin{equation}}
\def\ee{\end{equation}}
\newcommand{\beq}{\begin{eqnarray}}
\newcommand{\eeq}{\end{eqnarray}}

\def\ba{\begin{align}}
\def\ea{\end{align}}

\begin{document}
\title{Distinguishing black holes from horizonless objects\\
 through the excitation of resonances during inspiral}

\author{Vitor Cardoso$^{1,2}$,
Adri\'an del R\'{\i}o$^{1}$,
Masashi Kimura$^{3}$
}
\affiliation{${^1}$ CENTRA, Departamento de F\'{\i}sica, Instituto Superior T\'ecnico -- IST, Universidade de Lisboa -- UL,
Avenida Rovisco Pais 1, 1049 Lisboa, Portugal}
\affiliation{${^2}$ Theoretical Physics Department, CERN 1 Esplanade des Particules, Geneva 23, CH-1211, Switzerland}
\affiliation{${^3}$ Department of Physics, Rikkyo University, Tokyo 171-8501, Japan}
\begin{abstract}
How well is the vacuum Kerr geometry a good description of the dark, compact objects in our universe?
Precision measurements of accreting matter in the deep infrared and gravitational-wave measurements of
coalescing objects are finally providing answers to this question. Here, we study the possibility of resonant excitation of 
the modes of a central object -- taken to be very compact but horizonless -- during an extreme-mass-ratio inspiral. We show that for very compact objects resonances are indeed excited. However, we find that the impact of such excitation on the phase of
the gravitational-wave signal is negligible, since resonances are crossed very quickly during inspiral.
\end{abstract}

\preprint{RUP-19-20}

\maketitle

%\tableofcontents

%%%%%%%%%%%%%%%%%%%%%%%%%%%%%%%%%%%%%%%%%%%%%%%%%%%%
\section{Introduction}
%%%%%%%%%%%%%%%%%%%%%%%%%%%%%%%%%%%%%%%%%%%%%%%%%%%%
A remarkable feature of classical General Relativity is that vacuum spacetime can be curled to the extreme point of producing horizons, 
the boundaries of causally disconnected regions of spacetime that cloak singularities from far away observers. Such extraordinary property requires strong  observational evidence for black holes (BHs), a quest that should be placed alongside tests of the equivalence principle. In fact, dark compact horizonless objects are predicted to arise, at a phenomenological level, either when quantum effects are included or when beyond-the-standard model of particle physics is considered~\cite{Cardoso:2019rvt,Baibhav:2019rsa}.

Thus far, tests of the BH nature of compact objects which are based on gravitational-wave (GW) observations rely on i. small corrections to the GW phase as two compact bodies inspiral, driven by different multipolar structure, tidal deformation and heating~\cite{Hughes:2001jr,Krishnendu:2017shb,Cardoso:2017cfl,Maselli:2017cmm,Sennett:2017etc,Krishnendu:2018nqa,Datta:2019euh}; ii. echoes of the merger stage, induced by the presence of structure close to the gravitational radius of the final object~\cite{Cardoso:2016rao,Cardoso:2016oxy,Cardoso:2017cqb,Abedi:2016hgu,Conklin:2017lwb,Tsang:2018uie,Lo:2018sep,Nielsen:2018lkf,Uchikata:2019frs,Tsang:2019}. For a review see Ref.~\cite{Cardoso:2019rvt}.

Here, we study instead the possibility that the proper modes of oscillation of compact objects are excited and play a role in the inspiralling process. Previous studies focused on a special class of solutions -- boson stars -- which have a well defined underlying theory and are of interest from a particle-physics point of view. Resonant excitation of modes was found to be possible~\cite{Macedo:2013qea}. However, such self-gravitating solutions are never as compact as to be able to mimic the ringdown stage of BHs~\cite{Cardoso:2019rvt}. Therefore, here we turn to a (artificial) model describing the physics of objects whose surface sits deep down the gravitational potential~\footnote{Ultracompact objects -- so-called gravastars -- were investigated in Ref.~\cite{Pani:2010em}; it was shown that resonances can be excited during inspiral, but a proper detectability analysis was not performed.}. The compact object is assumed to be spherically symmetric. The exterior is vacuum and therefore described by the Schwarzschild geometry, down to the (hard) surface at
\be
r_0=2M(1+\epsilon)\,.
\ee
We consider both a toy model where a  particle coupled to a scalar field orbits the compact object, and a more realistic extreme-mass-ratio inspiral driven by GW emission.

%%%%%%%%%%%%%%%%%%%%%%%%%%%%%%%%%%%%%%%%%%%%%%%%%%%%
\section{Setup and results} 
%%%%%%%%%%%%%%%%%%%%%%%%%%%%%%%%%%%%%%%%%%%%%%%%%%%%

A pointlike  mass  $m_0$ coupled to a  scalar field with strength $\gamma$  orbits the central mass $M$ on a circular geodesic of radius $r_p\gg M$, emitting scalar and gravitational waves. We consider  linearized   scalar ($s=0$) and gravitational ($s=2$) field perturbations, and expand them in Fourier modes of frequency $\omega$. The angular dependence of these fields is  separated using spin-$s$ spherical harmonics, labeled by an angular number $\ell$ and an azimuthal number $m$.  We used matched asymptotic expansions to solve the relevant radial wave equations describing the linearized scalar and gravitational  field perturbations excited by the particle. 
 The technical details are relegated to appendices.
We are mostly interested in the possibility of excitation of the internal degrees of freedom of the massive object that could take place when the orbital frequency of the particle approaches a characteristic frequency of the system, and on the possibility that such excitations show up in the detected GW signal.  Such excitations of characteristic modes do not take place when the central object is a black hole since 
the modes of BHs are localized at the light ring, where timelike motion is general unstable~\cite{Berti:2009kk,Cardoso:2019rvt}; consequently, mode excitation by inspiralling bodies can be useful a priori in testing the nature of dark compact objects. 
We will assume, for simplicity, Dirichlet boundary conditions on the relevant master variables at the surface of the compact object. Gravitational fluctuations will most likely {\it not} interact significantly with any putative surface, but would cross unimpeded towards the center of the star, where they would be reflected. Thus, these artificial conditions are expected to mimic the physics we want to study.

%%%%%%%%%%%%%%%%%%%%%%%%%%%%%%%%%%%%%%%%%%%%%%%%%%%%
\subsection{Resonant frequencies}
%%%%%%%%%%%%%%%%%%%%%%%%%%%%%%%%%%%%%%%%%%%%%%%%%%%%
%{\color{red}We expand the linearized scalar and gravitational fields in Fourier modes of frequency $\omega$.}
Assuming reflective boundary conditions on the central object surface, the solution of the relevant wave equations appears as a linear combination of ingoing and outgoing waves at spatial infinity. Given the dissipative character of the system (i.e. energy escaping out to infinity), the mode frequencies $\omega$ are in general complex, and we shall denote its real and imaginary parts as $\omega_R$ and $\omega_I$, i.e., the characteristic modes are $\omega=\omega_R+i\omega_I$. When the solution is imposed to be a  purely outgoing wave at spatial infinity, corresponding to a maximum of energy flux emitted  by the system (i.e. a resonance), we get a condition on the frequencies. This condition defines the quasi-normal modes  $\omega$ of the system.

%and purely outgoing waves at spatial infinity leads in general to complex frequencies due to the dissipative character of the problem.
Define $\varpi=2M\omega$. In the small $\epsilon$ regime, we find resonant or quasinormal frequencies  at ($n=1,2,...$)
\beq
\varpi_{s=0}&\approx & \frac{n\pi}{\log \epsilon}- i \frac{(2 n \pi)^{2\ell+2} \Gamma(\ell+1)^6 }{ {4} |\log \epsilon|^{2\ell+3} \Gamma(2\ell+1)^2 \Gamma(2\ell+2)^2} \,,\nonumber\\
\varpi_{s=2}& \approx& \frac{n\pi}{\log \epsilon}- i \frac{(2 n \pi)^{2\ell+2} \Gamma(\ell+1)^2\Gamma(\ell-1)^2\Gamma(\ell+3)^2 }{ {8} |\log \epsilon|^{2\ell+3}  \Gamma(2\ell+1)^2\Gamma(2\ell+2)^2} \,,\nonumber
\eeq 
where recall that $\ell$ refers to the spherical harmonic mode. 
These results agree with previous analytical and numerical studies~\cite{Maggio:2018ivz,Cardoso:2019rvt}.

Now, to excite such quasinormal modes, the angular orbital frequency $\Omega$ of the particle needs to be tuned, $\omega_R=m\Omega$.
Thus, resonances occur when the radius of the circular orbit is $r_p = M^{1/3}/|\omega_R/m|^{2/3}$ (see \eqref{kepler})
with $r_0\omega_R \simeq n \pi/\log \epsilon$. For the orbit to be stable, the  radius should be $r>6M$, which implies that we
focus on 
\be
\epsilon<e^{-6 \sqrt{6}n\pi/m}\,.\label{min_eps}
\ee
%

%%%%%%%%%%%%%%%%%%%%%%%%%%%%%%%%%%%%%%%%%%%%%%%%%%%%
\subsection{Fluxes on and off resonance}
%%%%%%%%%%%%%%%%%%%%%%%%%%%%%%%%%%%%%%%%%%%%%%%%%%%%
In the absence of resonance with the central object, a particle on a circular orbit of radius $r_p$ gives rise to an energy flux whose dominant component is
\beq
\dot{E}_{s=0}&=&\frac{\gamma^2m_0^2M^2 }{12\pi\,r_p^4}\,,\label{scalar_flux_total}\\
\dot{E}_{s=2}&=&\frac{32}{5}\frac{m_0^2 M^3}{r_p^5}\,.\label{flux_BH} %= \frac{32}{5} \left(\frac{m_0}{M}\right)^2 (M \Omega)^{10/3}
\eeq
These fluxes correspond to the dominant dipolar and quadrupolar modes of the field for $s=0,2$ respectively
and agree with known expressions in the literature (most notably Einstein's quadrupole formula)~\cite{Cardoso:2007uy,Cardoso:2011xi,Yunes:2011aa,Poisson:1994yf}. The structure of the central object
is irrelevant in this regime (and thus a central BH would give rise to identical fluxes)~\cite{Poisson:1994yf}.

However, when the orbital frequency approaches a resonance frequency, the flux has a sharp peak equal to
%
%\be
%\dot{E}=-\frac{2^{2(l-5)/3}\Gamma(2l+2)}{\pi^{2(l-2)/3}\Gamma(-l+1/2)^2\Gamma(l+1)^2}\left(\log\epsilon\right)^{2(l-2)/3}\,,
%\ee
%
\begin{widetext}
\beq
%\dot{E}_{s=0}&\approx& \frac{\gamma^2 m_0^2 m^{4/3(\ell+1)}}{M^2 n^{2/3(\ell-2)}}\frac{2^{2(\ell-5)/3}\Gamma(2\ell+2)   \left(\log\epsilon\right)^{2(\ell-2)/3} }{\pi^{2(\ell-2)/3}\Gamma(-\ell+1/2)^2\Gamma(\ell+1)^2}\,,\\
 \dot{E}_{s=0} & \approx &  \frac{\gamma^2 m_0^2 m^{4 \ell/3}}{M^2 2^{14\ell/3+2} }  \frac{\Gamma(2\ell+2)^3\Gamma(2\ell+1)^4 (\gamma_{EM}+\psi(\ell+1))^2  \left(\log\epsilon\right)^{10\ell/3} }{(n\pi)^{10\ell/3} \Gamma(\ell+1)^{12}}\, ,  \\
\dot E_{s=2}& \approx&  \frac{m_0^2}{M^2} \frac{m^{4\ell/3} }{(n \pi)^{10\ell/3}}(\log \epsilon)^{10\ell/3} \frac{2\pi      \Gamma(2\ell+2)^3\Gamma(2\ell+1)^2 (\ell+1)^4 [2\gamma_{EM}+\psi(\ell-1)+\psi(\ell+3)]^2}{2^{2\ell/3} \Gamma(-\ell+1/2)^2  \Gamma(\ell+3)^5\Gamma(\ell-1)^3 \Gamma(\ell+1)^2 }\,,
\eeq
\end{widetext}
where $\psi(x)=\frac{1}{\Gamma(x)}\frac{d\Gamma(x)}{dx}$ is the digamma function, and $\gamma_{EM}$ is the Euler-Mascheroni constant.
These resonances have a radial width in orbital frequency of $\delta\Omega \sim \omega_I$, see \eqref{freqwidthw}. Notice that for most parameters of interest and for the dominant modes 
$\dot E$ at resonance is indeed larger than off-resonance. 

%%%%%%%%%%%%%%%%%%%%%%%%%%%%%%%%%%%%%%%%%%%%%%%%%%%%
\subsection{Impact of resonances on EMRIs}
%%%%%%%%%%%%%%%%%%%%%%%%%%%%%%%%%%%%%%%%%%%%%%%%%%%%

The pace at which inspiral proceeds is determined -- within a quasi-adiabatic approach -- by energy conservation. An increased flux at resonance implies that the inspiral towards an exotic horizonless objects proceeds faster, when compared to BH binaries. In turn, this effect might lead to an observable dephasing in gravitational waveforms. On the other hand, these are very narrow resonances and thus the accumulated energy release may be small enough that the effect is negligible.

To estimate the impact on the GW phase, one can compute the number of cycles accumulated during the 
resonant stage. If we denote by $f_i$ the orbital frequency of the particle at which a resonance starts, and by $f_f$  the orbital frequency at which the corresponding resonance finishes, then $f_f-f_i\sim \omega_I/2\pi$ is the resonance width and
\be
N_{\rm res}\sim \int_{f_i}^{f_f} \frac{f}{\dot{f}}df\,,
\ee
gives the desired number of cycles. Note that for gravitational waves, $f$ is also half the frequency of the quadrupolar waves being emitted. The variation of the frequency can be computed from the orbital parameters in a quasi-adiabatic fashion,
\be
\dot{f}\sim -\frac{3f}{2r}\left(\frac{dE_{\rm orb}}{dr}\right)^{-1}\dot{E}\,.
\ee
Here 
\be
E_{\rm orb}=m_0\frac{1-2M/r}{\sqrt{1-3M/r}}\,,
\ee
is the gravitational binding energy of the small point particle.

The number of cycles should be compared to those in a BH vacuum spacetime, $N_{\rm BH}$, obtained by using the flux~\eqref{flux_BH} in the previous expressions (we ignore fluxes through the horizon, since these are subdominant~\cite{Poisson:1994yf}). We find, for $l=m=2$ and for the dominant fundamental $n=1$ mode
\beq
%N_{\rm res}&\sim& 4.6\times 10^{-11}\,\frac{10^{-6}}{q}\left(\frac{10}{|\log\epsilon|}\right)^{43/3}\,,\\
N_{\rm res}&\sim& 2.4\times 10^{-11}\,\frac{10^{-6}}{q}\left(\frac{10}{|\log\epsilon|}\right)^{43/3}\,,\\
%
%\frac{N_{\rm res}}{N_{\rm BH}}&\sim& 1.9\times 10^{-10} \left(\frac{10}{|\log{\epsilon}|}\right)^{10}\,.
\frac{N_{\rm res}}{N_{\rm BH}}&\sim& 9.5\times 10^{-11} \left(\frac{10}{|\log{\epsilon}|}\right)^{10}\,,
\eeq
where $q\equiv m_0/M$ is the mass ratio.
%{\color{blue}[Adrian: for $\ell=2,m=2$ what I get is
%\beq
%N_{\rm res}&\sim& 6\times 10^{-10}\,\frac{M}{n\, m_0}\left(\frac{n\pi}{|\log\epsilon|}\right)^{43/3}\,,\\
%\frac{N_{\rm res}}{N_{\rm BH}}&\sim& 2\times 10^{-4} \left(\frac{n\pi}{|\log{\epsilon}|}\right)^{10}\,.
%\eeq Then, taking $n=1$ and multiplying the numerical factors by $(\pi/10)^{43/3}$, $(\pi/10)^{10}$, I get similar amplitudes, although slightly different. I assume $q\equiv m_0/M$ in formulas above.]}
%
Thus, the small object passes through resonances without any noticeable effect on the GW output. Higher modes are necessarily suppressed even further, since Eq.~\eqref{min_eps} forces $\epsilon$ to decrease exponentially with $n$.

It is, in principle, possible that the number of cycles spent in resonance is small, yet the signal is observable. However, the time $\delta t$ that it takes to cross the resonance can be estimated using $\delta t\sim \omega_I/\dot{\Omega}_{\rm orb}$, with
$\dot{\Omega}_{\rm orb}=d\Omega/dr\,dr/dE_{\rm orb}\dot{E}$. We find
\be
%\frac{\delta t}{M}\sim 9.4\times 10^{-10}\,\frac{10^{-6}}{q}\left(\frac{10}{|\log\epsilon|}\right)^{40/3}\,.
\delta t\sim 2.5\times 10^{-8}\,\frac{M}{10^6M_{\odot}}\,\frac{10^{-6}}{q}\left(\frac{10}{|\log\epsilon|}\right)^{40/3}\,{\rm s}\,.
\ee
This corresponds to a high-frequency ``glitch'', inaccessible by current or planned GW detectors.
%{\color{blue}[Adrian:  I get
%\be
%\frac{\delta t}{M}\sim \frac{3.8\times 10^{-9} }{n}\,\frac{M}{m_0}\left(\frac{n\pi}{|\log\epsilon|}\right)^{40/3}
%\ee
%For $n=1$, multiplying the numerical factor by $(\pi/10)^{40/3}$ reproduces  the same order of magnitude for the amplitude, $\sim 10^{-16}$.]
%}

%
%%%%%%%%%%%%%%%%%%%%%%%%%%%%%%%%%%%%%%%%%%%%%%%%%%%%
\subsection{Discussion}
%%%%%%%%%%%%%%%%%%%%%%%%%%%%%%%%%%%%%%%%%%%%%%%%%%%%
In conclusion, the inspiral of a small pointlike particle
around an ultracompact object can excite the characteristic modes of the central object, which carry important information on the nature of the latter.
However, our results indicate that such excitation does not have a significant impact of the phase 
of the GW, and leads to only a very high-frequency glitch. Thus, despite initial expectancies, resonant excitation of modes during inspiral turns out not to be a promising mechanism to help constraining the nature of dark, ultracompact objects. Our results are based on a simple-minded model for the supermassive object, it would certainly be desirable to extend the analysis to other self-gravitating objects whose surface lies extremely close to the Schwarzschild radius.

%%%%%%%%%%%%%%%%%%%%%%%%%%%%%%%%%%%%%%%%%%%%%%%%%%%%%%%%%%%%%%%%%%%%%%%%%%%%%
\noindent{\bf{\em Acknowledgments.}}
%%%%%%%%%%%%%%%%%%%%%%%%%%%%%%%%%%%%%%%%%%%%%%%%%%%%%%%%%%%%%%%%%%%%%%%%%%%%%
%
We are indebted to Francisco Duque for a careful reading of the manuscript, and for 
pointing out a typo in the scalar flux formula in the published version of the work.
We acknowledge financial support provided under the European Union's H2020 ERC 
Consolidator Grant ``Matter and strong-field gravity: New frontiers in Einstein's 
theory'' grant agreement no. MaGRaTh--646597.
This project has received funding from the European Union's Horizon 2020 research and innovation programme under the Marie Sklodowska-Curie grant agreement No 690904.
We acknowledge financial support provided by FCT/Portugal through grant PTDC/MAT-APL/30043/2017.
We acknowledge the SDSC Comet and TACC Stampede2 clusters through NSF-XSEDE Award Nos. PHY-090003.
The authors would like to acknowledge networking support by the GWverse COST Action 
CA16104, ``Black holes, gravitational waves and fundamental physics.''
%
% \end{acknowledgments}
%%%%%%%%%%%%%%%%%%%%%%%%%%%%%%%%%%%%%%%%%%%%%%%%%%%%%%%%%%%%%%%%%%%%%%%%%%%%%

\appendix

%%%%%%%%%%%%%%%%%%%%%%%%%%%%%%%%%%%%%%%%%%%%%%%%%%%%
\section{Scalar case}
\subsection{The setup}
%%%%%%%%%%%%%%%%%%%%%%%%%%%%%%%%%%%%%%%%%%%%%%%%%%%%
We start with a very simple toy problem, that of a massless scalar field $\Phi$ around a compact horizonless object of mass $M$ in a spacetime background of metric $g_{\mu\nu}$. The scalar field will be excited by introducing  a point-like particle of mass $m_0$ coupled to it and orbiting around the central object. The full dynamics is described by the action
\be
S[g,\Phi]=\int d^4x \sqrt{-g}\left(\frac{R[g]}{16\pi}-g^{\mu\nu}\partial_{\mu}\Phi\partial_{\nu}\Phi^*-2\gamma\Phi T\right)\,.
\ee
Here $R[g]$ denotes the Ricci scalar of the metric, $\gamma>0$ is a coupling constant, and $T$ is the trace of the stress tensor of the  particle.

We consider the point particle to be a small perturbation. Thus, the background spacetime is fixed and taken to be described by Schwarzschild exterior geometry, with coordinates $\{t,r,\theta,\phi\}$. All that remains is to solve the scalar field equation of motion coupled to the point-like particle:
\beq
%&&\frac{1}{k}\left(R^{\mu\nu}-\frac{1}{2}g^{\mu\nu}R\right)=\frac{1}{2}\left(\Phi^{*,\mu}\Phi^{,\nu}+\Phi^{\mu}\Phi^{*,\nu}\right)\nonum%ber\\
%
%&-&\frac{1}{2}g^{\mu\nu}\left(\Phi^*_{,\alpha}\Phi^{,\alpha}+2U\right)\,,\\
%
&&\frac{1}{\sqrt{-g}}\partial_{\mu}\left(\sqrt{-g}g^{\mu\nu}\partial_{\nu}\Phi\right)=\gamma T\,,\label{eq:KG}
\eeq
%
%%%%%%%%%%%%%%%%%%%%%%%%%%%%%%%%%%%%%%%%%%%%%%%%%%%%
%\subsection{Static backgrounds, real fields}
%%%%%%%%%%%%%%%%%%%%%%%%%%%%%%%%%%%%%%%%%%%%%%%%%%%%
For this matter we expand the fields in Fourier modes of frequency $\omega$ and in spherical harmonics $Y_{\ell m}$ as:
\beq
\Phi&=&\sum_{\ell,m}\int \frac{d\omega}{\sqrt{2\pi}}e^{-i\omega t}\frac{Z_{\ell m}(\omega,r)}{r}Y_{\ell m}(\theta,\phi)\,,\label{expansion_field}\\
r^2T&=&\sum_{\ell, m}\int \frac{d\omega}{\sqrt{2\pi}}e^{-i\omega t}T_{\ell m}(\omega)Y_{\ell m}(\theta,\phi) \label{defTlm}\, .
\eeq
where $\ell\geq 0$, $-\ell\leq m\leq \ell$. For static backgrounds, one finds the equation
\be
\frac{d^2Z_{\ell m}}{dr_*^2}+\left(\omega^2-f\left(\frac{\ell(\ell+1)}{r^2}+f'/r\right)\right)Z_{\ell m}=f\frac{\gamma}{r}T_{\ell m}\,,\label{wave_eq}
\ee
with $f=1-2M/r$, and $r^*$ denoting the tortoise coordinate. The object has a surface at $r=r_0$, or in tortoise coordinates $r_*=r_*^0$,
where we shall impose reflective conditions.
%%%%%%%%%%%%%%%%%%%%%%%%%%%%%%%%%%%%%%%%%%%%%%%%%%%%
\subsection{The source}
%%%%%%%%%%%%%%%%%%%%%%%%%%%%%%%%%%%%%%%%%%%%%%%%%%%%
%{\color{blue}We consider time-independent source terms $F(\Phi_0)$, but the results are easy to generalize. (A: What is $F$?) }

If $\tau$ denotes the proper time of the point particle along the world line $z^\mu(\tau)=(T(\tau),R(\tau),\vartheta(\tau),\varphi(\tau))$, the corresponding stress-energy tensor is given by
\beq
T^{\mu\nu}(x)&=&m_0\int_{-\infty}^{+\infty}\delta^{(4)}(x-z(\tau))\frac{dz^\mu}{d\tau}\frac{dz^\nu}{d\tau}d\tau\\
&=&m_0\frac{dt}{d\tau}\frac{dz^\mu}{dt}\frac{dz^\nu}{dt}\frac{\delta(r-R(t))}{r^2}\delta^{(2)}(\Omega-\Omega(t))\,, \nonumber
\eeq
where  the definition of the Dirac delta is taken as $\int\int\int\int\delta^{(4)}(x)\sqrt{-g}d^4x=1$.
We shall consider a stable circular  geodesic taking place in the equatorial plane. The particle will have  an orbit of radius $r=r_p>6M$ and an orbital frequency given by Kepler's law,
\be
\Omega^2=\frac{M}{r_p^3} \label{kepler}\,.
\ee
Then
\be
 g^{\mu\nu}T_{\mu\nu}(x)=\frac{-m_0}{\sqrt{-g} U^t}\delta(r-r_p)\delta(\theta-\pi/2)\delta(\phi-\Omega t)\,, \label{Taux}
\ee
where $U^t(r)=(1-3M/r)^{-1/2}$.

We can now  solve  the multipolar moments $T_{\ell m}(\omega)$. Equating (\ref{defTlm}) with (\ref{Taux}),
%Using (\ref{defTlm}) we  can derive now the multipolar moments $T_{\ell m}$
%
%\beq
%&&-\frac{m_0}{\sin\vartheta U^t}\delta(r-r_p)\delta(\vartheta-\pi/2)\delta(\varphi-\Omega t)\nonumber\\
%&=&\sum_{\ell m}\int \frac{d\omega}{\sqrt{2\pi}}e^{-i \omega t}T_{\ell m}(\omega)Y_{\ell m}(\vartheta,\varphi)\,.
%\eeq
%
multiplying both sides by $e^{i\omega' t}Y_{\ell'm'}^{*}$ and integrating on the sphere and in time, we get\footnote{We take the normalization of the spherical harmonics as  $\int_{\mathbb S^2} d\Omega \,  Y_{lm} Y^*_{\ell'm'}=\delta_{l'l}\delta_{m'm}$. We use the convention $\delta(x-x_0)=\frac{1}{2\pi}\int e^{it(x-x_0)}dt$.} 
%
%\be
%-\frac{m_0 Y_{\ell'm'}^{*}(\pi/2)e^{-im'\Omega t}}{U^t}\delta(r-r_p)=\int \frac{d\omega }{\sqrt{2\pi}}e^{-i\omega t}T_{\ell'm'}(\omega)\,.
%\ee
%
%{\color{blue} Note that for real frequencies the spherical harmonics are orthogonal on the sphere.} 
%
%Finally, multiplying both sides by $e^{i\omega' t}$ and integrating in time we find\footnote{We use the convention $\delta(x-x_0)=\frac{1}{2\pi}\int e^{it(x-x_0)}dt$.}
%
%\beq
%&&-\int dt \frac{m_0 Y_{\ell m}^{*}(\pi/2)e^{-i(m\Omega-\omega') t}}{U^t}\delta(r-r_p)\nonumber\\
%
%&&=\int dt \frac{d\omega}{\sqrt{2\pi}} e^{-i(\omega-\omega') t}T_{\ell m}(\omega)\,.
%\eeq
%
%Since
%
%\be
%\delta(x-x_0)=\frac{1}{2\pi}\int e^{it(x-x_0)}dt\,,\label{eq:delta}
%\ee
%
%we eventually find
%
\be
T_{\ell m}(\omega)=-\sqrt{2\pi}\frac{m_0 Y_{\ell m}^{*}(\pi/2)}{U^t}\delta(r-r_p)\delta(m\Omega-\omega)\,.
\ee
%

%%%%%%%%%%%%%%%%%%%%%%%%%%%%%%%%%%%%%%%%%%%%%%%%%%%%
\subsection{The formal solution}
%%%%%%%%%%%%%%%%%%%%%%%%%%%%%%%%%%%%%%%%%%%%%%%%%%%%
Define two independent solutions of the homogeneous ODE \eqref{wave_eq}  as
\beq
Z_1&\sim& e^{-i\omega(r_*-r_*^0)}-e^{i\omega(r_*-r_*^0)}\,,r_*\to r_*^0\,,\\
&\sim& A_{\rm in}e^{-i\omega r_*}+A_{\rm out}e^{i\omega r_*}\,,r_*\to +\infty \label{15}\\
Z_{2}&\sim& e^{i\omega r_*}\,,r_*\to +\infty
\eeq
The former one is  considered to have reflective boundary conditions on the central object surface $r_*^0$, while the latter describes purely outgoing waves at spatial infinity. The two of them are found  to be linearly independent by computing their Wronskian, which gives $2i\omega A_{in}$. The Green's function reads
\beq
G(r,r')=\frac{\theta(r'-r)Z_{2}(r')Z_1(r)+\theta(r-r')Z_{2}(r)Z_1(r')}{2i\omega A_{in}} \nonumber
\eeq
Then it is easy to show that, at large distances, the inhomogeneous solution is
\beq
Z_{\ell m}&=&e^{i\omega r_*}Z^\infty_{\ell m} \delta(m\Omega-\omega) \label{Zinfty}\\
&\equiv& -e^{i\omega r_*}\frac{\sqrt{2\pi}\,m_0 Y_{\ell m}^{*}(\pi/2)}{r_p U^t}\frac{\gamma Z_1(r_p)}{2i\omega A_{\rm in}}\delta(m\Omega-\omega) \,.\nonumber
\eeq
%
%%%%%%%%%%%%%%%%%%%%%%%%%%%%%%%%%%%%%%%%%%%%%%%%%%%%%%%%
\subsection{Energy flux}
%%%%%%%%%%%%%%%%%%%%%%%%%%%%%%%%%%%%%%%%%%%%%%%%%%%%%%%%
The energy flux emitted  to infinity by the scalar field is determined by 
\be
\dot E_{s=0}\equiv \frac{dE_{s=0}}{dt}=\lim_{r\to \infty}\int_{\mathbb S^2} d\theta d\phi \sqrt{-g}T_{rt}\, ,
\ee
where $T_{rt}$ is the relevant component of the  scalar  stress-energy tensor,
\be
T_{\mu\nu}= \nabla_\mu\Phi\nabla_\nu  \Phi^*-\frac{1}{2}g_{\mu\nu}\nabla_{\alpha}\Phi \nabla^{\alpha} \Phi^*   \, .
\ee
Taking into account the expansion in spherical harmonics \eqref{expansion_field} and  the asymptotic behavior 
$Z_{\ell m} \sim Z^\infty_{\ell m} e^{i\omega r } \delta(m\Omega-\omega)$ at large radial distances, we find
\begin{widetext}
\be
\frac{d^2E_{s=0}}{dtd\Omega}= \frac{1}{2\pi} \sum_{\ell' m'}\sum_{\ell m}\int d\omega d\omega' \omega\omega' \,Y_{\ell m}Y_{\ell'm'}e^{i(\omega-\omega')(r-t)}Z_\infty Z'_\infty\,\delta(m'\Omega-\omega')\delta(m\Omega-\omega)\,.
\ee
\end{widetext}
The normalization condition for the spherical harmonics reduce this to
\be
\dot E_{s=0}=\frac{1}{2\pi} \sum_{\ell m} \left(m\Omega\right)^2|Z^\infty_{\ell m}|^2 \label{flux}\, .
\ee
%

%%%%%%%%%%%%%%%%%%%%%%%%%%%%%%%%%%%%%%%%%%%%%%%%%%%%%%%%
\subsection{Matched asymptotic expansions}
%%%%%%%%%%%%%%%%%%%%%%%%%%%%%%%%%%%%%%%%%%%%%%%%%%%%%%%%
We now want to have an analytical understanding of the solutions of the homogeneous equation
at small frequencies. The homogeneous equation can be written in two equivalent forms 
\beq
&&\frac{d^2Z}{dr_*^2}+\left[\omega^2-f\left(\frac{\ell(\ell+1)}{r^2}+f'/r\right)\right]Z=0\,, \label{scalaraux1}\\
&&\left(r^2f(Z/r)'\right)'+\left[\frac{r^2}{f}\omega^2-\,\ell(\ell+1)\right](Z/r)=0\,, \label{scalaraux2}
\eeq
where primes stand for radial derivatives. %\footnote{from here on we drop multipolar indices $\ell$, $m$ where necessary in order to alleviate the notation.}

%%%%%%%%%%%%%%%%%%%%%%%%%%%%%%%%%%%%%%%%%%%%%%%%%%%%%%%%
\subsubsection{The near-region solution}
%%%%%%%%%%%%%%%%%%%%%%%%%%%%%%%%%%%%%%%%%%%%%%%%%%%%%%%%

We follow the procedure in Refs.~\cite{Cardoso:2004nk,Maggio:2018ivz}. Consider first a ``near-region''
where $r-r_0\ll 1/\omega$ (we assume $r_0\sim 2M$ to good approximation). Then, the second equation above can be written as
\be
x\left(\frac{Z}{r}\right)''+\left(\frac{Z}{r}\right)'+\left(\frac{\varpi^2}{x}-\frac{\ell(\ell+1)}{(1-x)^2}\right)\left(\frac{Z}{r}\right)=0\,,  \label{aux1}
\ee
where now primes are derivatives with respect to $x\equiv f$, and we introduced the dimensionaless frequency $\varpi\equiv \omega r_H$. 
Notice that at the object surface $r=r_0$, $x_0=1-r_H/r_0$, where $r_H=2M$. For $r_0=r_H(1+\epsilon)$, then $x_0\sim \epsilon$.

Defining now $Z/r=x^{i\varpi}(1-x)^{\ell+\epsilon+1}F$, and neglecting $O(\varpi^2)$ and $O(\epsilon)$ terms in the coefficient of $F$, one finds the standard hypergeometric equation,
\be
x(1-x)\partial^2_xF+\left(c-(a+b+1)x\right)\partial_xF-abF \approx 0\,,
\ee
with
\be
a=\ell+\epsilon+ 1+i2\varpi\,,\quad b=\ell+\epsilon+1\,,\quad c=1+i2\varpi\,. \label{scalarparameters}
\ee
Given that $c$ is not an integer, around $x=0$ two linearly independent solutions are $_{2}F_{1}(a, b,c,x)$ and $x^{1-c}\,_{2}F_{1}(a-c+1,b-c+1,2-c,x)$. The general solution in the near-region is then
\beq
Z&=&Ax^{-i\varpi}(1-x)^{\ell+\epsilon}F(a-c+1,b-c+1,2-c,x)\nonumber\\
&+&Bx^{i\varpi}(1-x)^{\ell+\epsilon}F(a,b,c,x)\,.
\eeq
For BHs one imposes boundary conditions corresponding to purely ingoing waves at the horizon, and that implies $B=0$. For ``exotic compact objects'' (ECOs) with Dirichlet BCs at the surface $x\sim 0$,    $Z=Ax_0^{-i\varpi}+Bx_0^{i\varpi}$, thus $B=-Ax_0^{-2i\varpi}$.

To understand the far-region behavior of the above solution we use the transformation properties of hypergeometric functions,
\beq
&&\-F(a,b,c,z)=\-\frac{\Gamma(c)\Gamma(c-a-b)}{\Gamma(c-a)\Gamma(c-b)}F(a,b,a+b-c+1,1-z) \label{hypergeometricproperties}\\
&&+\-\-(1-z)^{c-a-b}\frac{\Gamma(c)\Gamma(a+b-c)}{\Gamma(a)\Gamma(b)}F(c-a,c-b,c-a-b+1,1-z)\,.  \nonumber
\eeq
Therefore, the large $r$ behavior is
\begin{widetext}
\beq
Z&\sim&  \frac{r_H^\ell\Gamma(-1-2\ell-2\epsilon)}{r^{\ell}\Gamma(-\ell-\epsilon)}\left[\frac{A\, \Gamma(1-2i\varpi)}{\Gamma(-\ell-\epsilon-2i\varpi)}+\frac{B\, \Gamma(1+2i\varpi)}{\Gamma(-\ell-\epsilon+2i\varpi)}\right] \nonumber\\
&+&\frac{r^{\ell+1} \Gamma(1+2\ell+2\epsilon)}{r_H^{\ell+1}\Gamma(\ell+\epsilon+1)} \left[\frac{B\, \Gamma(1+2i\varpi)}{\Gamma(\ell+\epsilon+1+2i\varpi)}+\frac{A\, \Gamma(1-2i\varpi)}{\Gamma(\ell+\epsilon+1-2i\varpi)}\right]  \label{nearregion}\,.
\eeq
\end{widetext}
%%%%%%%%%%%%%%%%%%%%%%%%%%%%%%%%%%%%%%%%%%%%%%%%%%%%%%%%
\subsubsection{The far-region solution}
%%%%%%%%%%%%%%%%%%%%%%%%%%%%%%%%%%%%%%%%%%%%%%%%%%%%%%%%

In the far-region {(i.e. when $r\gg r_0$)}, the wave equation reduces to
\be
\partial^2_r Z+\left(\omega^2-\ell(\ell+1)/r^2\right)Z=0\,,
\ee
with solutions
\be
Z=r^{1/2}\left(\alpha J_{\ell+1/2}(\omega r)+\beta J_{-\ell-1/2}(\omega r)\right)\,,\label{eq_Bessel}
\ee
which for small $r$ reduces to
\be
Z\sim \alpha \frac{\omega^{\ell+1/2}r^{\ell+1}}{2^{\ell+1/2}\Gamma(\ell+3/2)}+\beta \frac{2^{\ell+1/2}r^{-\ell} }{\omega^{\ell+1/2}\Gamma[-\ell+1/2]}\label{farregion}\,.
\ee
%

%%%%%%%%%%%%%%%%%%%%%%%%%%%%%%%%%%%%%%%%%%%%%%%%%%%%%%%%
\subsubsection{Matching}
%%%%%%%%%%%%%%%%%%%%%%%%%%%%%%%%%%%%%%%%%%%%%%%%%%%%%%%%
From the behavior of the Bessel functions at $r\to \infty$, the solution (\ref{eq_Bessel}) is asymptotic to
\beq
Z\sim e^{i(\omega r+\frac{\pi}{2}\ell)} \sqrt{\frac{2}{\pi\omega}} \frac{\beta-i \alpha (-1)^{\ell}}{2} + (i\to -i)\,,\nonumber
\eeq
and by demanding now the behaviour (\ref{15}) we  get
\be
A_{\rm in}=\frac{e^{-i\ell\pi/2}\left(\beta+i\alpha e^{i\ell\pi}\right)}{\sqrt{2\pi\omega}}\,.\label{Ain_expression}
\ee
Now we want to express $\alpha$ and $\beta$ in terms of the boundary parameters $A$, $B$. In order to do this we proceed to match the small $r$ behaviour of the far-region solution to the large $r$ behaviour of the near-region solution.

%For any $\ell\in \mathbb R$ such that $2\ell+1\neq 0,1,2, \dots$ and $\ell\neq 0,1,2, \dots$,
%
%\beq
%\Gamma(-2\ell-1) & = &   \frac{\pi}{\sin[2(\ell+1)\pi]\Gamma(2\ell+2)}=\frac{\pi}{\sin(2\ell \pi)\Gamma(2\ell+2)}\nonumber \\
%\Gamma(-\ell)  & = &  \frac{\pi}{-\sin(\ell \pi) \Gamma(\ell+1)}\,.
%\eeq
%
%Then
%
%\be
%\frac{\Gamma(-2\ell-1)}{\Gamma(-\ell)}=\frac{-\Gamma(1+\ell)}{2\cos(\pi\,\ell)\Gamma(2\ell+2)}\,.
%\ee
%
%The RHS is well-defined also for positive integer values of $\ell$, so we define the LHS for any $\ell \in \mathbb R$ by analytic continuation. In particular, for integer values of $\ell$ the formula above yields
%
First of all, we use the result
\beq
\frac{\Gamma(-2\ell-1-2\epsilon)}{\Gamma(-\ell-\epsilon)}= \frac{(-1)^{\ell+1} \Gamma(\ell+1)}{2\Gamma(2\ell+2)}+O(\epsilon)\,.\label{gammaidentity}
\eeq
On the other hand, we can also expand the Gamma function to find
\be
\frac{\Gamma(\ell-\epsilon-2i\varpi)}{\Gamma(-\ell-\epsilon-2i\varpi)}=\frac{(\epsilon+2i\varpi)(-1)^{\ell+1}}{\ell-\epsilon-2i\varpi}\, \prod_{k=1}^\ell (k^2+(\epsilon+2i \varpi)^2)\,. \label{gammaidentity2}
\ee
Then, in the limit in which $\varpi\ll1$, we get
\beq
\alpha 
&\approx &\frac{2^{\ell+1/2}(A+B)\Gamma(\ell+3/2)\Gamma(2\ell+1)}{\varpi^{\ell+1/2}\sqrt{r_H}\Gamma(\ell+1)^2}\,, \label{scalaralfa}\\
\beta&\approx & % \frac{i2^{-1/2-l}(A-B)\varpi^{l+3/2}l! \Gamma(-l+1/2)\Gamma(l+1)^2}{l\sqrt{r_H}(2l+1)!\Gamma(l)}
\frac{i(A-B)\varpi^{\ell+3/2}\ell! \Gamma(-\ell+1/2) \Gamma(\ell+1)}{ 2^{1/2+\ell}\sqrt{r_H}(2\ell+1)!} \label{scalarbeta}
\eeq
Finally, we can write
%
%\beq
%A_{\rm in}=\frac{(A-B)4^\ell\varpi^{2\ell+2}(\ell!)^6+(A+B)((2\ell)!)^2((2\ell+1)!)^2}
%{(\ell!)^3(2\ell)!(2\ell+1)! (-2i\varpi)^{\ell+1}}\,. \label{AinAB}
%\eeq
%
\begin{widetext}
\beq
A_{\rm in}\approx i\, \frac{\varpi^{2\ell+2}(A-B)\Gamma(-\ell+1/2)\Gamma(\ell+1)^5 {2^{-1}}+e^{i\ell\pi}\ell\sqrt{\pi}(A+B)\Gamma(2\ell)\Gamma(2\ell+2)^2}{(2i)^{\ell}\varpi^{\ell+1}\sqrt{\pi}\Gamma(\ell+1)^3\Gamma(2\ell+2)}\,. \label{AinAB}
\eeq
\end{widetext}

%%%%%%%%%%%%%%%%%%%%%%%%%%%%%%%%%%%%%%%%%%%%%%%%%%%%%%%%
\subsection{The black hole flux formulas}
%%%%%%%%%%%%%%%%%%%%%%%%%%%%%%%%%%%%%%%%%%%%%%%%%%%%%%%%
As mentioned before, for BHs $B=0$. If we focus on the small-frequency regime $\varpi\ll 1$, the second term in \eqref{eq_Bessel} is suppressed against the first, and ~\cite{Abramowitz:1970as}
\be
Z(r_p)\approx  \alpha \sqrt{r_p}J_{\ell+1/2}(\omega r_p)\sim \alpha \sqrt{r_p}\frac{(\omega r_p)^{\ell+1/2}}{2^{\ell+1/2}\Gamma(\ell+3/2)}\,.\nonumber
\ee
From (\ref{flux}) and (\ref{Zinfty}) we find, 
\beq
\dot{E}_{s=0} \approx \sum_{\ell m}\frac{m_0^2\gamma^2|Y_{\ell m}(\frac{\pi}{2},0)|^2}{4 r_p^2|A_{\rm in}|^2}|Z_1(r_p)|^2\,, \label{scalarfluxadv}
\eeq
where we used $\omega=m\Omega$.

We can explicitly write $Y_{\ell m}(\pi/2,0)$ using
\beq
&&Y_{\ell m}(\pi/2,0)=\sqrt{(\ell+ m)!(\ell- m)!}\nonumber\\
&\times&\frac{(-1)^{(\ell+ m)/2}(1 + (-1)^{\ell+ m})\sqrt{1 + 2\ell}}{2^{2 +\ell} \sqrt{\pi}\,((\ell+ m)/2)!((\ell- m)/2)!}\,.
\label{spherical_equator}
\eeq
For definiteness, focus on the $m = \ell$ modes, for which $Y_{\ell\ell}(\pi/2,0)=(-1)^\ell/(2^\ell \ell!)\sqrt{(2\ell+1)!/(4\pi)}$.
Using now the expressions for $A_{\rm in}$ and $\alpha$, in the limit in which $\varpi \ll 1$, 
\beq
\dot E_{s=0}\approx \frac{m_0^2\gamma^2}{4\pi} \frac{\omega^{2\ell+2}r_p^{2\ell}}{\Gamma(2l+2)}\,.
\eeq
Using Kepler's law (\ref{kepler}),  $\omega=\ell \Omega$,  this can also be expressed as
\beq
\dot E_{s=0}\approx  \frac{m_0^2\gamma^2}{4\pi} \frac{\ell^{2\ell+2} M^{\ell+1}}{r_p^{\ell+3}\Gamma(2\ell+2)}\,.
\eeq
In particular, for $\ell=m=1$ one finds
\beq
\dot{E}^{l=m=1}_{s=0}\approx \frac{\gamma^2m_0^2M^2}{24 \pi\,r_p^4} \label{VitorBH}\,.
\eeq
This result agrees very well with full numerical solutions of the inhomogeneous wave equation (at a radius $r_p=50M$ the simple analytical formula yields the correct result with 8\% accuracy). Taking into account also the $l=-m=1$ mode, one finds the flux formula Eq.~\eqref{scalar_flux_total} as the leading order term in a post-Newtonian expansion. Such expression is in perfect agreement with results in the literature~\cite{Cardoso:2007uy,Cardoso:2011xi,Yunes:2011aa}.

%%%%%%%%%%%%%%%%%%%%%%%%%%%%%%%%%%%%%%%%%%%%%%%%%%%%%%%%
\subsection{The QNMs of ECOs}
%%%%%%%%%%%%%%%%%%%%%%%%%%%%%%%%%%%%%%%%%%%%%%%%%%%%%%%%
%
%
\begin{table}[t]
\begin{tabular}{ccc}
\hline
\hline
$\ell$    & Approximate formula & Numerical calculation \\
\hline
\hline
 0 & $0.237 - 7.486\times 10^{-3}i$ & $0.220 - 8.646\times 10^{-3}i$\\
 1 & $0.237- 1.075\times 10^{-5}i$ & $0.260 - 4.113\times 10^{-5}i$\\
 2 & $0.237- 2.471\times 10^{-9}i$ & $0.284 - 2.590\times 10^{-8}i$\\
\hline
\hline
\end{tabular}
\caption{
The value of QNM frequencies $\omega_{\rm QNM}\times r_H$ for a scalar field and for an ultracompact object with $\epsilon = 10^{-6}$. 
}
\label{tab:qnm_frequencies}
\end{table}
For ECOs we have $B=-Ax_0^{-2i\varpi}$. The energy flux is determined by  (\ref{flux}) with (\ref{Zinfty}). We  expect resonances then at the poles of (\ref{AinAB}). These are the quasi-normal mode (QNM) frequencies of these objects~\cite{Cardoso:2019rvt,Maggio:2018ivz}.

We compute the QNM frequencies  as the roots of $A_{\rm in}(\varpi) = 0$. By setting (\ref{AinAB}) equal to zero we can write
\be
e^{i2 \varpi \log x_0}R(\varpi)=1\,,\quad R(\varpi)=\frac{1-F_{s=0}(\ell)\varpi^{2\ell+2}}{1+F_{s=0}(\ell)\varpi^{2\ell+2}}\,,\nonumber
\ee 
with
\beq
F_{s=0}(\ell)&=& \frac{\Gamma(1/2-\ell)\Gamma(\ell+1)^5}{{2} \ell \Gamma(2\ell)\Gamma(2\ell+2)^2\sqrt{\pi}(-1)^{\ell+1}}  \,.\label{scalarF}
\eeq
We proceed  to solve the equation iteratively, by solving $e^{i2\varpi_{i+1} \log x_0}R(\varpi_i)=1$ for $\varpi_{i+1}$, with the initial input $ \varpi_0=0$. In the first iteration one obtains
\beq
\varpi_{1}=\frac{n\pi}{|\log x_0|}\,, \label{eq:QNM}
\eeq
for all $n \in \mathbb Z$. The second iteration gives  (since $x_0<1$ the logarithm is negative)
\beq
\varpi_{2}=\frac{n\pi}{|\log x_0|}-i \frac{1}{2|\log x_0|} \log\left[ \frac{1-F_{s=0}(\ell) \left(\frac{n\pi}{\log x_0} \right)^{2\ell+2} }{1 + F_{s=0}(\ell) \left(\frac{n\pi}{\log x_0} \right)^{2\ell+2} }\right] \,.\nonumber
\eeq 
As argued in the main text $x_0\ll e^{-6 \sqrt{6}n\pi/m}$, so $|\log x_0|\gg n\pi$, and we can approximate
\beq
\varpi_{2}\approx \frac{n\pi}{|\log x_0|}+ i \frac{F_{s=0}(\ell)(n \pi)^{2\ell+2} }{ |\log x_0|^{2\ell+3}}\,. \label{widthaux}
\eeq 
$F_{s=0}(\ell)$ can be further simplified and then 
%
%\beq
%\varpi_{2}& \approx &  \frac{n\pi}{\log x_0}- i \frac{(2 n \pi)^{2\ell+2} \Gamma(\ell+1)^6 }{ {4} |\log x_0|^{2\ell+3} (2\ell+1)^2 \Gamma(2\ell+1)^4} \,.
%\eeq 
%
the next-to-leading order result for the QNMs is
\be
\varpi_{s=0} \simeq \frac{n\pi}{|\log x_0|} - i \frac{(2 n \pi)^{2\ell+2} (\ell!)^6}{    {4} (1+2\ell)^2 ((2\ell)!)^4}
\frac{1}{|\log x_0|^{3+2\ell}} \, .\label{QNMfreqS}
\ee
The presence of a non-vanishing imaginary part with the correct sign accounts for the exponential decay of the mode in time (stability). This analytical prediction works reasonably well, as checked with the numerical implementation (see Table I).

%%%%%%%%%%%%%%%%%%%%%%%%%%%%%%%%%%%%%%%%%%%%%%%%%%%%%%%%
\subsection{The ECO flux formulas}
%%%%%%%%%%%%%%%%%%%%%%%%%%%%%%%%%%%%%%%%%%%%%%%%%%%%%%%%
For $B=-Ax_0^{-2i\varpi}$ (\ref{AinAB}) gives
\begin{widetext}
\be
%A_{\rm in}=iA\,  \frac{\varpi^{2\ell+2}(1+x_0^{-2i\varpi})\Gamma(-\ell+1/2)\Gamma(\ell+1)^5 {2^{-1}}+e^{i\ell\pi}\ell\sqrt{\pi}(1-x_0^{-2i\varpi})\Gamma(2\ell)\Gamma(2\ell+2)^2}{(2i)^{\ell} \varpi^{\ell+1}\sqrt{\pi}\Gamma(\ell+1)^3\Gamma(2\ell+2)}\,. 
|A_{\rm in}|^2\approx  |A|^2\frac{\varpi^{4\ell+4} \cos^2(\varpi\log x_0 )\Gamma(1/2-\ell)^2\Gamma(\ell+1)^{10}  +4\ell^2\pi \sin^2(\varpi \log x_0) \Gamma(2\ell)^2\Gamma(2\ell+2)^4}{4^\ell \varpi^{2\ell+2}\pi \Gamma(\ell+1)^6  \Gamma(2\ell+2)^2 }\, . \label{Ain}
\ee
\end{widetext}
where we neglected the imaginary part of $\varpi$ to write $|1+x_0^{2i\varpi}|^2=2+2\cos(2\varpi \log x_0)=4\cos^2(\varpi \log x_0)$ and $|-1+x_0^{2i\varpi}|^2=2-2\cos(2\varpi \log x_0)=4\sin^2(\varpi \log x_0)$.

 On the other hand, notice from (\ref{scalaralfa}) that  the leading order behaviour of  $\alpha$ is $ \sim (1-x_0^{-2i\varpi})$, but around QNM frequencies this is highly suppressed\footnote{It is proportional to ${\rm Im} \varpi \log x_0$, and according to (\ref{QNMfreqS}) it is tiny.}. Hence we need to go back to the exact expression for $\alpha$,
\beq
\alpha & = & \frac{2^{\ell+1/2}\Gamma(\ell+3/2)\Gamma(2\ell+1)}{\varpi^{\ell+1/2}\sqrt{r_H}\Gamma(\ell+1)^2}\nonumber\\
&\times&\left[\frac{B\, \Gamma(1+2i\varpi)}{\Gamma(\ell+1+2i\varpi)}+\frac{A\, \Gamma(1-2i\varpi)}{\Gamma(\ell+1-2i\varpi)}\right]\,, \nonumber
\eeq
and work out the  subsequent leading order behaviour. This  will dominate  $Z_1(r_p)$, in contrast to the BH case.
Doing this, and neglecting the $\beta$ contribution again:
\beq
&&|Z_1(r_p)|^2  \approx   |\alpha|^2 \frac{\omega^{2\ell+1}r_p^{2\ell+2}}{2^{2\ell+1}\Gamma(\ell+3/2)^2} \\
 && =  \left(\frac{r_p}{r_H}\right)^{2\ell+2} \frac{16 |A|^2 \cos^2(\varpi \log x_0 )\varpi^{2}\Gamma(2\ell+1)^2(\gamma_{EM}+\psi(\ell+1))^2}{\Gamma(\ell+1)^4} \nonumber
\eeq
where we introduced the function $\psi(x)=\Gamma'(x)/\Gamma(x)$ and the Euler-Mascheroni constant $\gamma_{EM}$. 
Using (\ref{scalarfluxadv}), near these QNM frequencies, the leading order flux reads
\begin{widetext}
\be
\dot{E}_{s=0}\approx \left(\frac{r_P}{r_H}\right)^{2\ell}\frac{\gamma^2m_0^2 \varpi^{2\ell+4}\cos^2{(\varpi\log{x_0})}(\gamma_{EM}+\psi(\ell+1))^2\Gamma(2\ell+1)^{2}\Gamma(2\ell+2)^3}{r_H^2  \left(\varpi^{4\ell+4}\cos^2{(\varpi\log{x_0})}\Gamma(-\ell+1/2)^2\Gamma(\ell+1)^{10}+{4} \ell^2\pi \Gamma(2\ell)^2\Gamma(2\ell+2)^4\sin^2{(\varpi\log{x_0})}\right)}\,. \label{fluxnearQNM}
\ee
\end{widetext}
At the QNM frequencies the flux is maximum and yields
%
%\be
%\dot{E}=-\frac{2^{2(l-5)/3}\Gamma(2l+2)}{\pi^{2(l-2)/3}\Gamma(-l+1/2)^2\Gamma(l+1)^2}\left(\log\epsilon\right)^{2(l-2)/3}\,,
%\ee
%
\beq
&&\dot{E}_{s=0}\approx \frac{\gamma^2 m_0^2 m^{4\ell/3}}{M^2 2^{14\ell/3+2} } \times \label{fluxpeak} \\ 
&&\frac{\Gamma(2\ell+2)^3\Gamma(2\ell+1)^4 (\gamma_{EM}+\psi(\ell+1))^2  \left(\log\epsilon\right)^{10\ell/3} }{(n\pi)^{10\ell/3} \Gamma(\ell+1)^{12}}\,, \nonumber
\eeq
where we used the orbital parameters to excite the resonance at \eqref{eq:QNM}.

%%%%%%%%%%%%%%%%%%%%%%%%%%%%%%%%%%%%%%%%%%%%%%%%%%%%%%%%
\subsection{Resonance widths}
%%%%%%%%%%%%%%%%%%%%%%%%%%%%%%%%%%%%%%%%%%%%%%%%%%%%%%%%
Given the Kepler relation (\ref{kepler}), the energy flux (\ref{fluxnearQNM})  is a function of the radial distance only. To find the width of the ECO resonances we look for a value $r$ of the radial distance that fulfills 
$
\dot E_{}(r) = \frac{1}{2} \dot E_{ }(r_{\rm QNM}) ,
$
where   $r_{\rm QNM}$ denotes  the  value that gives the resonant frequencies $\varpi_{\rm QNM}$ written in (\ref{eq:QNM}), i.e. $\varpi_{\rm QNM}^2 /m^2=4M^3/r_{\rm QNM}^3$. 

We assume that the frequency band will be small, so that we can expand the function $\varpi(r)/m=\sqrt{4M^3/r^3}$ in Taylor series at $r_{\rm QNM}$ as:
\be
\varpi(r) = \varpi_{\rm QNM}\left[1 -\frac{3}{2} \frac{r-r_{\rm QNM}}{r_{\rm QNM}} +\frac{15}{4} \left(\frac{r-r_{\rm QNM}}{r_{\rm QNM}} \right)^2\right]+\dots \label{bandw}
\ee
The scalar flux (\ref{fluxnearQNM}) can be written now as
\be
\dot E_{s=0}(r) = \frac{ a_{\ell m} \varpi^{2\ell/3+4}(r)\cos^2(\varpi(r)\log x_0)}{b_{\ell }\varpi^{4\ell+4}(r)\cos^2(\varpi(r)\log x_0)+c_{\ell } \sin^2(\varpi(r)\log x_0)}\,, \nonumber
\ee
with $b_{\ell}\equiv \Gamma(-\ell+1/2)^2\Gamma(\ell+1)^{10}>0$ and $c_{\ell}\equiv 4\ell^2\pi\Gamma(2\ell)^2\Gamma(2\ell+2)^{4}>0$ ($a_{\ell m}$ will not be needed). The flux peaks at $r_{\rm QNM}$, and gives $\dot E_{s=0}(r_{\rm QNM})=\frac{a_{\ell m}}{b_{\ell}} \varpi^{-10/3\ell}_{\rm QNM}$. By expanding the energy flux up to second order we get\footnote{That the coefficient in front of $r-r_{\rm QNM}$ is not identically zero is just a residue due to the approximations. In an exact approach this term would not appear. It does not play any role in what follows.}
\begin{widetext}
\beq
\frac{\dot E_{s=0}(r)}{\dot E_{s=0}(r_{\rm QNM})}=\left[1+5\ell \frac{r-r_{\rm QNM}}{r_{\rm QNM}} +\left(\frac{r-r_{\rm QNM}}{r_{\rm QNM}} \right)^2\left(\frac{-35\ell+50\ell^2}{4}-\frac{9c_{\ell} n^2 \pi^2 \varpi_{\rm QNM}^{-4-4\ell}}{4b_{\ell}} \right) \right]+O\left( \frac{r-r_{\rm QNM}}{r_{\rm QNM}} \right)^3 \,.\nonumber
\eeq
\end{widetext}
Demanding now $\dot E_{s=0}(r) = \frac{1}{2} \dot E_{s=0}(r_{\rm QNM}) $ one gets a quadratic equation for $(r-r_{\rm QNM})/r_{\rm QNM}\equiv y$,
\be
y^2\left[ \frac{-35\ell+50\ell^2}{4}-\frac{9c_{\ell} n^2 \pi^2 \varpi_{\rm QNM}^{-4-4\ell}}{4b_{\ell}} \right]+y(5\ell)+\frac{1}{2}=0 \,,\nonumber
\ee
which leads to two solutions, $r_{+}$ and $r_{-}$,  corresponding to the outer and inner radius of the width, respectively. The solution, in the approximation in which $\varpi_{\rm QNM}\ll 1$ (recall (\ref{QNMfreqS})), takes the simple form:
\beq
y_{\pm} \approx \pm \frac{\varpi_{\rm QNM}^{2\ell+2}}{n\pi} \sqrt{  \frac{2 b_{\ell} }{9c_{\ell}}    }\,.
\eeq
From Eq.~\eqref{scalarF} it can be readily checked that $\sqrt{\frac{ b_{\ell} }{c_{\ell}} }= F_{s=0}(\ell)(-1)^{\ell+1}$. Thus we finally arrive at
\beq
\left| \frac{r_+-r_-}{r_{\rm QNM}} \right| & \approx &  \sqrt{\frac{8}{9} } \frac{ {\rm Im}\varpi_{\rm QNM}}{ {\rm Re} \varpi_{\rm QNM}}  \label{radialwidthS} \\ & \approx&  \sqrt{\frac{8}{9} } \frac{(2 n \pi)^{2\ell+1} \Gamma(\ell+1)^6 }{ {2} |\log x_0|^{2\ell+2} (2\ell+1)^2 \Gamma(2\ell+1)^4} \,, \nonumber
\eeq
where we used the form (\ref{widthaux}).
%\begin{widetext}
The  width in frequency around the value $\varpi_{\rm QNM}$ can be inferred from (\ref{bandw}) and (\ref{radialwidthS}). It reads:
\beq
2|\varpi_{\rm QNM}-\varpi(r_+)| & \approx &  3 \varpi_{\rm QNM}\frac{|r_+-r_{\rm QNM} |}{r_{\rm QNM}}\nonumber \\
&\approx &  %\frac{3}{2} \varpi_{QNM}\left| \frac{r_+-r_-}{r_{QNM}} \right| \approx 
\sqrt{2} {\rm Im} \varpi_{\rm QNM} \,.\label{freqwidthw}
\eeq

%%%%%%%%%%%%%%%%%%%%%%%%%%%%%%%%
\section{Gravitational case}
%%%%%%%%%%%%%%%%%%%%%%%%%%%%%%%%%
\subsection{The setup}

We consider gravitational radiation within the framework of metric perturbations around the Schwarzschild geometry. These perturbations are excited by the inspiral of a small mass around the Schwarzschild BH. We will study these perturbations using the Newman-Penrose formalism combined with the Regge-Wheeler analysis, following Ref.~\cite{Poisson:1993vp}. 

The fundamental perturbation field is the Weyl scalar $\Psi_4$, which is well adapted to analyze outgoing GWs.  We expand it as %\footnote{This is (2.2) in [Poisson1992]. See chapter 4 of [Chandrasekhar], (236) and (247), and text around for more details. The  normalization factor $r^4$ accounts for the Peeling behaviour of $\Psi_4$.}
\beq
r^4\Psi_4&=&\sum_{\ell,m}\int_{-\infty}^{\infty} d\omega R_{\omega \ell m}(r)e^{-i\omega t}\ {_{-2}Y}_{\ell m}(\theta,\phi)\,,
\eeq
where ${_{-2}Y}_{\ell m}$ denote the spherical harmonics of spin-weight $s=-2$~\cite{Goldberg:1966uu}. The sums are restricted to $\ell \geq 2$  and $-\ell \leq m \leq \ell$. With this ansatz for the Weyl scalar, the linearised Newman-Penrose equations lead to %\footnote{This is (2.3) in [Poisson1992]. See  chapter 4 in [Chandrasekar], (266)-(267) to find the explicit derivation when $T_{\omega \ell m}=0$.}
\be
\left[ r^2f \frac{d^2}{dr^2} - 2(r-M)\frac{d}{dr}+U(r)  \right]R_{\omega \ell m}(r)=T_{\omega \ell m} \,,\,\label{Teukolski}
\ee
with 
\be
U(r)= f^{-1}[(\omega r)^2-4 i \omega(r-3M)]- (\ell-1)(\ell+2)\,,
\ee
and source term $T_{\omega \ell m}$ given in detail in (2.6) of \cite{Poisson:1993vp}. 

We employ the Green's function method to obtain a solution of the previous equation. To build the Green's function we need two linearly independent solutions of the homogeneous equation.	We take $R^H_{\omega \ell m}$ and $R^{\infty}_{\omega \ell m}$. The latter one will describe purely outgoing waves escaping to infinity. The former one shall denote purely ingoing waves in the BH case (corresponding to a perfect absorber), and purely reflected waves in the ECO case (corresponding to a perfect mirror). For the considered ODE the Wronskian of these two solutions must be a constant, and so it can be evaluated at any value of $r$. According the their definitions, these solutions should have the following asymptotic behaviours %\footnote{This appears after Eq. (2.6) in [Poisson1992]. See chapter 4 in [Chandrasekar], first line of (366) with the notation $Y\equiv B$, and text around, to understand the behaviour.}
\beq
%R^H_{\omega \ell m}(r)  & \sim &  r^4 f^2 \tilde A e^{-i\omega r*}+ \tilde B e^{i\omega r*} ... \, , \hspace{0.5cm} r\to 2M\\
R^H_{\omega \ell m}(r\to +\infty) &  \sim & \frac{1}{r} B_{\omega \ell m}^{in} e^{-i\omega r_*}+ B_{\omega \ell m}^{out} r^3 e^{i\omega r_*}  \, , \hspace{0.5cm}   \label{asymptoticconditions}\\
R^{\infty}_{\omega \ell m}(r\to +\infty) & \sim  &  r^3 e^{i\omega r_*} \, , 
\eeq
and the Wronskian yields, %\footnote{The normalization of the Wronskian is the one that makes the principal symbol of the differential operator to be 1. That is way we divide by $r^2f(r)$.}
\beq
\lim_{r\to \infty}\frac{ R^{H}_{\omega \ell m} R^{\infty \, '}_{\omega \ell m}-R^{H \, '}_{\omega \ell m} R^{\infty}_{\omega \ell m} }{r^2 f(r)} = 2 i \omega  B_{\omega \ell m}^{in}\,.
\eeq
Here, primes denote differentiation with respect to $r$. Thus, both solutions are linearly independent as long as $B_{\omega \ell m}^{in}\neq 0$ and $\omega \neq 0$. The Green's function is
\beq
G(r,r') & = & \frac{\left[ \theta(r'-r)\frac{R_{\infty}(r')R_H(r)}{r'^2 f(r')}+\theta(r-r')\frac{R_{\infty}(r)R_H(r')}{r'^2f(r')} \right]}{2 i \omega  B_{\omega \ell m}^{in}}\,.\nonumber
\eeq
Following the standard theory, we can write the inhomogeneous solution as
\beq
R_{\omega \ell m}(r ) & = & \int_{2M}^{\infty}dr' \frac{G(r,r') T_{\omega \ell m}(r')}{r'^2 f(r')} \\ 
&\sim& \frac{r^3 e^{i\omega r_*}}{2i\omega B_{\omega \ell m}^{in}}\int_{2M}^{\infty}dr' \frac{R^{H}_{\omega \ell m}(r') T_{\omega \ell m}(r')}{r'^4 f^2(r')} \nonumber
\eeq
where in the last step we considered the limit $r\to \infty$.

%%%%%%%%%%%%%%%%%%%%%%%%%
\subsection{Energy flux}
%%%%%%%%%%%%%%%%%%%%%%%%%
To calculate the energy flux it is helpful to introduce an auxiliary quantity $Z_{\ell m}$ by 
$R_{\omega \ell m}(r\to \infty ) = m_0 Z_{\ell m} \delta(\omega-m \Omega) r^3 e^{i\omega r_*}$. Then the flux formula reads
\beq
\dot E_{s=2}=m_0^2 \sum_{\ell m} \frac{ |Z_{\ell m}|^2}{4\pi \omega^2}\,.
\eeq
The auxiliary function can be calculated from the stress-energy tensor of a point particle orbiting around a Schwarzschild BH and the result is  \cite{Poisson:1993vp} 
\begin{widetext}
\beq
Z_{\ell m} & = & \frac{\pi}{i \omega r_p^2 B_{\omega \ell m}^{in}} \left[ \left(_0 b_{\ell m}+2i _{-1}b_{\ell m}\left[1+\frac{i\omega r_p}{2f(r_p)} \right]-i _{-2}b_{\ell m}\left[1-\frac{M}{r_p}+\frac{i \omega r_p}{2}\right] \right)R^H_{\omega \ell m}(r_p) \right. \nonumber\\
 &  &\hspace{2cm} \left. -\left( i _{-1}b_{\ell m}- _{-2}b_{\ell m}(1+\frac{i\omega r_p}{f(r_p)}) \right) r_p R^{H\, '}_{\omega \ell m}(r_p)-\frac{1}{2} \, _{-2}b_{\ell m} r_p^2  R^{H\, ''}_{\omega \ell m}(r_p) \right] \,,\label{Zlm}
\eeq
\end{widetext}
with coefficients
\beq
_{0}b_{\ell m} & = &\frac{\sqrt{(\ell-1)\ell(\ell+1)(\ell+2)}}{2 \sqrt{1-\frac{3M}{r_p}} } \, _{0}Y_{\ell m}\left(\frac{\pi}{2 },0 \right)  \,,\label{b0}\\
_{-1}b_{\ell m} & = & \sqrt{ \frac{M(\ell-1)(\ell+2)}{r_p -3M } }\, _{-1}Y_{\ell m}\left(\frac{\pi}{2},0 \right) \,, \label{b1} \\
_{-2}b_{\ell m} & = & \sqrt{\frac{M r_p^2}{r_p-3M}}\, _{-2}Y_{\ell m}\left(\frac{\pi}{2},0 \right) \Omega \,.\label{b2} 
\eeq
From the  identity $_{s}Y_{\ell -m}\left(\frac{\pi}{2},0 \right) = {_{s}Y}_{\ell m}\left(\frac{\pi}{2},0 \right)(-1)^{s+\ell}$ it is easy to see that $Z_{\ell -m}=(-1)^{\ell}\bar Z_{\ell m}$. Consequently, the final expression for the energy flux is
\beq
\dot E_{s=2}=m_0^2 \sum_{\ell =2}^{\infty} \sum_{m=1}^{\ell} \frac{ |Z_{\ell m}|^2}{2\pi \omega^2}\,. \label{fluxfinal}
\eeq
with $\omega=m\Omega$.
%%%%%%%%%%%%%%%%%%%%%%%%%%%%%%%%%%%%%%%%%%%%%%%%%%%%%%%%%%%%%%%%%%%%%%
\subsection{Reducing the problem to solving the Regge-Wheeler equation}
%%%%%%%%%%%%%%%%%%%%%%%%%%%%%%%%%%%%%%%%%%%%%%%%%%%%%%%%%%%%%%%%%%%%%%
Chandrasekhar \cite{Chandrasekhar:1985kt} showed that if $X_{\omega \ell m}(r)$ is a solution to the Regge-Wheeler equation%\footnote{For an explicit derivation see chapter 4 in [Chandrasekhar], leading to equations (27)-(28).}:
\beq
\left\{\frac{d^2}{dr_*^2}+\omega^2-V(r)\right\} X_{\omega \ell m}(r)=0\,.
\eeq
with effective potential
\beq
V(r)=f\left[ \frac{\ell(\ell+1)}{r^2}-\frac{6M}{r^3}\right]\,.
\eeq
then %\footnote{This is summarized in section IID, eq. (2.19) of [Poisson1992]. See chapter 4 in [Chandrasekhar]  equations (266), (288) and (302).} 
%\begin{widetext}
\beq
R_{\omega \ell m}(r)&=&r^3V(r)X_{\omega \ell m}(r)  \label{transformationR}\\
&+&r^3\left[\frac{2}{r}\left(1-\frac{3M}{r}\right)+2i\omega \right] \left[ \frac{\partial}{\partial r_* }+i\omega\right]X_{\omega \ell m}(r)  \nonumber
\eeq
%\end{widetext}
is a solution to the homogeneous equation (\ref{Teukolski}). So rather than working with the ODE derived from the Newman-Penrose formalism, it is more convenient to solve the Regge-Wheeler equation first and then apply the Chandrasekar transformation to obtain the Weyl component $R^H_{\omega \ell m}(r)$ that governs the energy flux (\ref{fluxfinal}).

The relevant solution $X^H_{\omega \ell m}(r)$ for our problem has the following asymptotic conditions, inherited from (\ref{asymptoticconditions}):
\beq
X^H_{\omega \ell m}(r\to 2M)  & \sim &   A e^{-i\omega r_*}+  B e^{i\omega r_*}  \, ,  \\
X^H_{\omega \ell m}(r\to +\infty) &  \sim & A_{\omega \ell m}^{in} e^{-i\omega r_*}+ A_{\omega \ell m}^{out} e^{i\omega r_*}  \, ,  \label{asymptoticG}
\eeq
where \cite{Chandrasekhar:1985kt, Poisson:1993vp} %\footnote{the transformation rules are explained in section 32, chapter 4 of [Chandrasekhar]. More precisely, compare () and (), } 
\beq
 B_{\omega \ell m}^{in} & = &\frac{ \left[ 12i M\omega-(\ell-1)\ell(\ell+1)(\ell+2) \right]  }{4\omega^2} A_{\omega \ell m}^{in}\,.  \label{transformationB}   \\
B_{\omega \ell m}^{out} & = & -4\omega^2 A_{\omega \ell m}^{out} \,.
\eeq
As in the scalar case, for BHs we shall fix $B=0$, while for ECOs we have $B=-Ax_0^{-2i\omega}$.
%%%%%%%%%%%%%%%%%%%%%%%%%%%%%%%%%%%
\subsection{Some simplifications}
%%%%%%%%%%%%%%%%%%%%%%%%%%%%%%%%%%%
Although  equation (\ref{Zlm}) seems complicated at first, it can be further simplified.
Recall that we  consider the problem of having the particle far away from the BH, so that $r_p\gg M$. Using the Kepler Law $M\omega/m=M \Omega=(M/r_p)^{3/2}$ this automatically implies that $\varpi=2M\omega \ll 1$. Inspection of (\ref{b0})-(\ref{b2}) shows that 
\beq
_0 b_{\ell m}  & \sim  & O(\varpi^0)\,,\\
_{-1} b_{\ell m}  & \sim  & O(\varpi^{1/3})\,,\\
_{-2} b_{\ell m}  & \sim  & O(\varpi^{2/3})\,.
\eeq
This means that $Z_{\ell m}$ in (\ref{Zlm}) is dominated by the contribution involving $_0 b_{\ell m}$, unless $ _{0}Y_{\ell m}\left(\frac{\pi}{2},0 \right) $ vanishes, which according to \eqref{spherical_equator} happens  when $\ell+m$ is an odd number. When this happens, $Z_{\ell m}$ is dominated by the term involving  $_{-1} b_{\ell m}$, but this is suppressed in our approximation. For a given $\ell$ therefore, the energy flux (\ref{fluxfinal}) will be dominated by modes for which $\ell+m$ is even. The leading order expression for $Z_{\ell m}$ is then
\beq
Z_{\ell m} & \approx &  \frac{\pi}{i \omega r_p^2 B_{\omega \ell m}^{in}}  {_0 b}_{\ell m}R^H_{\omega \ell m}(r_p)\,. \label{Zimproved}
\eeq

All that remains now is to calculate $A_{in}(\omega)$ and $X^H_{\omega \ell m}(r)$. From (\ref{transformationB}) and (\ref{transformationR}),  we get  $B_{in}(\omega)$ and $R^H_{\omega \ell m}$, respectively. Then finally we will use (\ref{Zimproved}) to calculate (\ref{fluxfinal}). For the purpose of calculating $A_{in}(\omega)$ and $X^H_{\omega \ell m}(r)$ we shall follow closely the strategy done in the scalar case.
%%%%%%%%%%%%%%%%%%%%%%%%%%%%%%%%%%%%%%%%%
\subsection{Matched asymptotic expansions}
%%%%%%%%%%%%%%%%%%%%%%%%%%%%%%%%%%%%%%%%%%
The homogeneous Regge-Wheeler equation can be written in two equivalent forms
\beq
&&\frac{d^2X}{dr_*^2}+\left(\omega^2-f\left(\frac{\ell(\ell+1)}{r^2}+(1-{s^2}) f'/r\right)\right)X=0\,, \label{RW1}\\
&&\left(r^2f\left[\frac{X}{r}\right]'\right)'+\left(\frac{r^2}{f}\omega^2-\,\ell(\ell+1)+{s^2}r f'\right)\frac{X}{r}=0\,, \hspace{1cm}\label{RW2}
\eeq
where primes stand for radial derivatives, and $s=2$. The scalar case is recovered by taking $s=0$.
%%%%%%%%%%%%%%%%%%%%%%%%%%%%%%%%%%%%
\subsubsection{Near-region solution}
%%%%%%%%%%%%%%%%%%%%%%%%%%%%%%%%%%%%
We analyze first the regime in which $\omega \ll \frac{1}{r-2M}$. Comparing (\ref{RW2}) with (\ref{scalaraux2}) it is clear that we can recycle the results done for the scalar case, as long as we keep track of the new factor $s^2$. Equation (\ref{aux1}) reads now
%\begin{widetext}
\be
x\left(\frac{X}{r}\right)''+\left(\frac{X}{r} \right)'+\left[\frac{\varpi^2}{x (1-x)^4}-\frac{\ell(\ell+1)}{(1-x)^2}+\frac{s^2}{1-x}\right] \frac{X}{r}=0\,,  \nonumber
\ee
where here primes are derivatives with respect to $x\equiv f$ and $\varpi\equiv \omega r_H$. Making the substitution $X/r=x^{i\varpi} (1-x)^{\ell+\epsilon+1} F$ we find, after neglecting $O(\varpi^2)$ and $O(\epsilon)$ terms,
\beq
x(1-x)\partial^2_xF+\left(c-(a+b+1)x\right)\partial_xF-(ab-s^2)F=0\,, \nonumber
%+F\frac{\varpi^2}{x}\left[\frac{1}{(1-x)^3}-1+x \right]+{\color{red}s^2}F=0\,,
\eeq
%\end{widetext}
with $a,b,c$ the  parameters (\ref{scalarparameters}). Now we want to reabsorbe the $s^2$ term into these parameters.
Define $a'$ and $b'$ by $ab-s^2=a'b'$ and $a+b=a'+b'$. Solving this system leads to
\beq
a' & = & % \ell+\epsilon+1+i \varpi+2\sqrt{1-\varpi^2/4} =
\ell+\epsilon+3+i \varpi+O(\varpi^2)\,, \nonumber\\
b' & = & %\ell+\epsilon+1+i\varpi-2\sqrt{1-\varpi^2/4}=
\ell+\epsilon-1+i \varpi+O(\varpi^2)\,.\nonumber
\eeq
%
%By neglecting  $O(\varpi^2)$ contributions one ends up with an hypergeometric differential equation, just as in the scalar case. 
The general solution in the near-region is then
\beq
X_{\omega \ell m}&=&Ax^{-i\varpi}(1-x)^{\ell+\epsilon}F(a'-c+1,b'-c+1,2-c,x)\nonumber\\
&+&Bx^{i\varpi}(1-x)^{\ell+\epsilon}F(a',b',c,x) \label{sol2F1G}\,.
\eeq
To study the far-region behavior of the above solution use again the transformation properties of hypergeometric functions (\ref{hypergeometricproperties}). Then, we find the large $r$ behavior ($x \sim 1$)
\begin{widetext}
\beq
X_{\omega \ell m}&\sim& \left[\frac{r_H}{r}\right]^{\ell} \Gamma(-1-2\ell-2\epsilon)    \left(\frac{B\Gamma(1+2i\varpi)}{\Gamma(-\ell-\epsilon+2+i\varpi)\Gamma(-\ell-\epsilon-2+i\varpi)} +\frac{A\Gamma(1-2i\varpi)}{\Gamma(-\ell-\epsilon+2-i\varpi)\Gamma(-\ell-\epsilon-2-i\varpi)} \right) \nonumber\\
&+& \left[\frac{r}{r_H}\right]^{\ell+1}\Gamma(1+2\ell+2\epsilon) \left( \frac{B\Gamma(1+2i\varpi)}{\Gamma(\ell+\epsilon+3+i\varpi)\Gamma(\ell+\epsilon-1+i\varpi)} +\frac{A\Gamma(1-2i\varpi)}{\Gamma(\ell+\epsilon+3-i\varpi)\Gamma(\ell+\epsilon-1-i\varpi)} \right) \label{nearregionG}\,.
\eeq
\end{widetext}

%%%%%%%%%%%%%%%%%%%%%%%%%%%%%%%%%%%%%%%%%%%%%
\subsubsection{The far-region solution}
%%%%%%%%%%%%%%%%%%%%%%%%%%%%%%%%%%%%%%%%%%%%%%%%%%%%%%%%

In the far-region {(i.e. when $r\gg r_0$)}, the wave equation (\ref{RW1}) reduces to
\be
\partial^2_r X+\left(\omega^2-\ell(\ell+1)/r^2\right)X=0\,,
\ee
with solutions
\be
X_{\omega \ell m}=r^{1/2}\left(\alpha J_{\ell+1/2}(\omega r)+\beta J_{-\ell-1/2}(\omega r)\right)\,,\label{eq_BesselG}
\ee
and for small $r$ reduces to
\be
X_{\omega \ell m}\sim \alpha \frac{\omega^{\ell+1/2}}{2^{\ell+1/2}\Gamma(\ell+3/2)}r^{\ell+1}+\beta \frac{2^{\ell+1/2}}{\omega^{\ell+1/2}\Gamma[-\ell+1/2]}r^{-\ell} \label{farregionG}\,.
\ee
%

%%%%%%%%%%%%%%%%%%%%%%%%%%%%%%%%%%%%%%%%%%%%%%%%%%%%%%%%
\subsubsection{Matching}
%%%%%%%%%%%%%%%%%%%%%%%%%%%%%%%%%%%%%%%%%%%%%%%%%%%%%%%%

From the behavior of the Bessel functions {at $r\to \infty$, the solution (\ref{eq_BesselG}) is asymptotic to
\beq
X_{\omega \ell m}\sim e^{i(\omega r+\frac{\pi}{2}\ell)} \sqrt{\frac{2}{\pi\omega}} \frac{\beta-i \alpha (-1)^{\ell}}{2} +(i\to -i)\,.
\eeq
and by demanding now the behaviour (\ref{asymptoticG})} we  get
\be
A^{\rm in}_{\omega\ell m}=\frac{e^{-i\ell\pi/2}\left(\beta+i\alpha e^{i\ell\pi}\right)}{\sqrt{2\pi\omega}}\,.\label{Ain_expressionG}
\ee
{Now we want to express $\alpha$ and $\beta$ in terms of the boundary parameters $A$, $B$. As in the scalar case,  we proceed to match the small $r$ behaviour of the far-region solution to the large $r$ behaviour of the near-region solution.} 

First, we use the approximation (correct up to ${\cal O}(\varpi,\epsilon)$)
\beq
\frac{\Gamma(-2\ell-2\epsilon-1)}{\Gamma(2-\ell-\epsilon+i\varpi)}& = & 
 \frac{(-1)^{\ell+1}\Gamma(\ell-1)}{2\Gamma(2\ell+2)}\,.\label{AuxG1}
\eeq
In addition, we can use
\beq
&& \frac{\Gamma(1+2i\varpi)}{\Gamma(-\ell-\epsilon+i \varpi)}= \frac{\Gamma(1+2i\varpi)(i \varpi-\epsilon)(-1)^{\ell}}{\Gamma(-\ell+i \varpi)\Gamma(\ell-\epsilon+\varpi i+1)} \nonumber\\
&\times&\prod_{k=1}^{\ell} [k^2-(i\varpi-\epsilon)^2] \,,
\eeq
to show that, up to ${\cal O}(\varpi^2,\epsilon)$,
\be
\frac{\Gamma(1+2i\varpi)}{\Gamma(-2-\ell-\epsilon+i \varpi)} = (-1)^{\ell} i\varpi \Gamma(\ell+3) \,. \label{AuxG2}
\ee

Using the above results, we find, in the limit $\varpi \ll 1$, 
\beq
\alpha & \approx & \frac{2^{\ell+1/2}}{\sqrt{r_H}\varpi^{\ell+1/2}}\frac{\Gamma(1+2\ell)\Gamma(\ell+3/2)}{\Gamma(\ell+3)\Gamma(\ell-1)}(A+B)\,,\nonumber\\
\beta &  \approx & \frac{\varpi^{\ell+3/2}\Gamma(-\ell+1/2)\Gamma(\ell+3) \Gamma(\ell-1)i}{2^{\ell+3/2}\sqrt{r_H} \Gamma(2\ell+2)}(A-B)\,.\nonumber
\eeq
Putting these expressions into (\ref{Ain_expressionG}) we find
\begin{widetext}
\beq
A^{in}_{\omega\ell m}=i\frac{(A+B)\Gamma(1+2\ell)4^{\ell+1}\Gamma(\ell+3/2)\Gamma(2\ell+2)(-1)^{\ell} +(A-B)\Gamma(-\ell+1/2) \varpi^{2(\ell+1)} \Gamma(\ell+3)^2\Gamma(\ell-1)^2  }{2^{\ell+2}i^{\ell}\sqrt{\pi}\varpi^{\ell+1}\Gamma(\ell+3)\Gamma(\ell-1) \Gamma(2\ell+2)}\,.  \label{AinG}
\eeq
\end{widetext}
%

%%%%%%%%%%%%%%%%%%%%%%%%%%%%%%%%%%%%%%%%%%%%%%%%%%%%%%%%
\subsection{The black hole flux formulas}
%%%%%%%%%%%%%%%%%%%%%%%%%%%%%%%%%%%%%%%%%%%%%%%%%%%%%%%%
For BHs, $B=0$. Equation (\ref{AinG}) together with (\ref{transformationB}) gives, in the approximation   $\varpi \ll 1$,
%\beq
%A_{in}(\omega) 
%& \approx & i A  \frac{\Gamma(1+2\ell)2^{\ell}\Gamma(\ell+3/2)(i)^{\ell}  }{ \sqrt{\pi}\varpi^{\ell+1}\Gamma(\ell+3)\Gamma(\ell-1)} 
%\eeq
%
\be
|B^{in}_{\omega \ell m}|^2
%\approx  \frac{|A_{in}(\omega)|^2}{16\omega^4}\frac{\Gamma(\ell+3)^2}{\Gamma(\ell-1)^2} 
\approx  |A|^2  \frac{\Gamma(1+2\ell)^2 4^{\ell-2}\Gamma(\ell+3/2)^2 }{ \pi  \omega^4 \varpi^{2(\ell+1)}\Gamma(\ell-1)^4} \,.
\ee
From (\ref{transformationR}), the Weyl scalar function can be calculated as
\be
R^H(r_p)\approx  r_p\ell(\ell+1) X(r_p)+2 r_p^2 \left(\frac{d}{d r}+i\omega \right)X|_{r_p}\,, \label{transformationapprox}
\ee
where we notice that $\varpi = \left(M/r_p\right)^{3/2}\ll M/r_p$. To calculate the derivative we use equation (\ref{sol2F1G}), and then we take the large distance limit with (\ref{hypergeometricproperties}). The large distance limit of $X_{\omega\ell m}$ is taken from  (\ref{nearregionG}). Doing the calculation in detail, and staying always in the approximation in which $\varpi \ll1$ (or equivalently $r_p\gg r_H$), we end up with
\beq
 X(r_p)  & \approx &   \left( \frac{r_p}{r_H} \right)^{\ell+1} A \frac{ \Gamma(1+2\ell)}{\Gamma(\ell+3)\Gamma(\ell-1)}\,,\\
\frac{d}{dr} X|_{r_p} & \approx &  \left( \frac{r_p}{r_H} \right)^{\ell+1} \frac{A}{r_p} \frac{(\ell+1) \Gamma(1+2\ell)}{\Gamma(\ell+3)\Gamma(\ell-1)}\,.
\eeq
Then it is straightforward to write from (\ref{transformationapprox})
\beq
R^H_{\omega \ell m}(r_p) & \approx &  \left( \frac{r_p}{r_H} \right)^{\ell+1} A r_p\frac{ \Gamma(1+2\ell)}{\Gamma(\ell+1)\Gamma(\ell-1)}\,.
\eeq
From (\ref{Zimproved}), (\ref{b0}) and using \eqref{spherical_equator} for odd $m+\ell$, we get, after several simplifications,
%\beq
%|Z_{\ell m}|^2 & \approx &  \frac{\pi^2}{ \omega^2 r_0^4 }  |R^H_{\omega \ell m}(r_0)|^2  \frac{ \varpi^{2(\ell+1)} \Gamma(\ell-1)^3 \Gamma(\ell+3)\Gamma(2\ell+2)  }{|A|^2 \Gamma(1+2\ell)^2 4^{2\ell+2}\Gamma(\ell+3/2)^2\Gamma(\ell+1)^2}     \approx  \frac{\pi^2}{  r_0^4 }    \frac{ \varpi^{2\ell} \Gamma(\ell-1)^3 \Gamma(\ell+3)\Gamma(2\ell+2)  }{\Gamma(1+2\ell)^2 4^{2\ell+2}\Gamma(\ell+3/2)^2\Gamma(\ell+1)^2}    \left( \frac{r_p}{r_H} \right)^{2(\ell-1)}
%\eeq
\beq
|Z_{\ell m}|^2 & \approx & \frac{ m^{2\ell+4}M^{\ell+2}}{ r_p^{\ell+6}}\frac{4 \pi\Gamma(\ell+3)\Gamma(\ell-1)}{\Gamma(2\ell+2)\Gamma(\ell+1)^2}\,,
%\frac{\pi^{3/2} m^{2\ell+4}M^{\ell+2}}{2^{2\ell-1}r_p^{\ell+6}}\frac{\Gamma(\ell+3)\Gamma(\ell-1)}{\Gamma(\ell+3/2)\Gamma(\ell+1)^3} 
\eeq
and the energy flux (\ref{fluxfinal}) is finally:
\beq
\dot E_{s=2} \approx m_0^2 \sum_{\ell =2}^{\infty} \sum_{m={\rm even}}^{\ell} \frac{ m^{2\ell+2}M^{\ell+1}}{ r_p^{\ell+3}}\frac{2 \Gamma(\ell+3)\Gamma(\ell-1)}{\Gamma(2\ell+2)\Gamma(\ell+1)^2}\,.\nonumber
% \frac{\pi^{1/2} m^{2\ell+2}M^{\ell+1} }{2^{2\ell} r_p^{\ell + 3}} \frac{\Gamma(\ell+3)\Gamma(\ell-1)}{\Gamma(\ell+3/2)\Gamma(\ell+1)^3} 
\eeq

Note that for the leading order contribution, $\ell=m=2$, we get
\beq
\dot E_{s=2}\approx \frac{32}{5}\frac{m_0^2 M^3}{r_p^5} = \frac{32}{5} \left(\frac{m_0}{M}\right)^2 (M \Omega)^{10/3}\,,
\eeq
which agrees with the well-known Einstein's quadrupole formula.

%%%%%%%%%%%%%%%%%%%%%%%%%%%%%%%%%%%%%%%%%%%%%%%%%%%%%%%%
\subsection{The QNMs of ECOs}
%%%%%%%%%%%%%%%%%%%%%%%%%%%%%%%%%%%%%%%%%%%%%%%%%%%%%%%%
Following the same reasoning as in the scalar case, we derive the QNM frequencies  as the roots of $A^{\rm in}_{\omega \ell m} = 0$. By setting (\ref{AinG}) equal to zero we can write
\be
e^{i2 \varpi \log x_0}R(\varpi)=1, \hspace{0.5cm} R(\varpi)=\frac{1-F_{s=2}(\ell)\varpi^{2\ell+2}}{1+F_{s=2}(\ell)\varpi^{2\ell+2}}\,,\nonumber
\ee 
now with
\be
F_{s=2}(\ell) =\frac{\Gamma(1/2-\ell)\Gamma(\ell+3)^2\Gamma(\ell-1)^2(-1)^{\ell+1}}{    {4}^{\ell+1}   \Gamma(\ell+3/2)\Gamma(2\ell+1)\Gamma(2\ell+2) } \,.\label{gravitationalF}
\ee
We solve the equation iteratively, by solving $e^{i2\varpi_{i+1} \log x_0}R(\varpi_i)=1$ for $\varpi_{i+1}$, with the initial condition $ \varpi_0=0$. The solution is given in (\ref{widthaux}), but with the new $F_{s=2}$.
% In the first iteration one gets
%\beq
%\varpi_{1}=\frac{n\pi}{\log x_0}, 
%\eeq
%for any $n \in \mathbb Z$. The second iteration yields
%\beq
%\varpi_{2}=\frac{n\pi}{\log x_0}-i \frac{1}{2|\log x_0|} \log\left[ \frac{1-F_{s=2}(\ell) \left(\frac{n\pi}{\log x_0} \right)^{2\ell+2} }{1 + F_{s=2}(\ell) \left(\frac{n\pi}{\log x_0} \right)^{2\ell+2} }\right] \nonumber
%\eeq 
%Since $|\log x_0|>> n\pi$, we can write
%\beq
%\varpi_{2}\approx \frac{n\pi}{\log x_0}+ i \frac{F(\ell)(n \pi)^{2\ell+2} }{ |\log x_0|^{2\ell+3}} \label{widthaux}
%\eeq 
%$F(\ell)_{s=2}$ can be further simplified and  
%\beq
%\varpi_{2}^{} & \approx  & \frac{n\pi}{\log x_0}- i \frac{(2 n \pi)^{2\ell+2} \Gamma(\ell+1)^2\Gamma(\ell-1)^2\Gamma(\ell+3)^2 }{ {8} |\log x_0|^{2\ell+3} (2\ell+1) \Gamma(2\ell+1)^3\Gamma(2\ell+2)} 
%\eeq 
After some simplifications, the result for the quasi-normal modes of gravitational perturbations is 
\be
\varpi_{s=2} \simeq \frac{n\pi}{\log x_0}- i \frac{(2 n \pi)^{2\ell+2} \Gamma(\ell+1)^2\Gamma(\ell-1)^2\Gamma(\ell+3)^2 }{ {8} |\log x_0|^{2\ell+3} (2\ell+1) \Gamma(2\ell+1)^3\Gamma(2\ell+2)}\,.  \label{QNMfreq}
\ee
Again, the non-vanishing imaginary part has the correct sign  and takes into account the exponential decay of the mode in time (stability).

%%%%%%%%%%%%%%%%%%%%%%%%%%%%%%%%%%%%%%%%%%%%%%%%%%%%%%%%
\subsection{The ECO flux formulas}
%%%%%%%%%%%%%%%%%%%%%%%%%%%%%%%%%%%%%%%%%%%%%%%%%%%%%%%%
For ECOs we have $B=-Ax_0^{-2i\varpi}$. Equation (\ref{AinG}) yields 
\begin{widetext}
%\beq
%A_{in}(\omega)=iA \frac{(1-x_0^{-2i\varpi})\Gamma(1+2\ell)4^{\ell+1}\Gamma(\ell+3/2)\Gamma(2\ell+2)(-1)^{\ell} +(1+x_0^{-2i\varpi})\Gamma(-\ell+1/2) \varpi^{2(\ell+1)} \Gamma(\ell+3)^2\Gamma(\ell-1)^2  }{2^{\ell+2}i^{\ell}\sqrt{\pi}\varpi^{\ell+1}\Gamma(\ell+3)\Gamma(\ell-1) \Gamma(2\ell+2)} \label{AinECOG}
%\eeq and 
\beq
|A_{in}(\omega)|^2 \approx |A|^2 \frac{\Gamma(1+2\ell)^2 4^{2\ell+2}\Gamma(\ell+3/2)^2\Gamma(2\ell+2)^2 \sin^2(\varpi \log x_0) + \Gamma(-\ell+1/2)^2 \varpi^{4(\ell+1)} \Gamma(\ell+3)^4\Gamma(\ell-1)^4 \cos^2(\varpi \log x_0) }{4^{\ell+1} \pi \varpi^{2(\ell+1)}\Gamma(\ell+3)^2\Gamma(\ell-1)^2\Gamma(2\ell+2)^2} \,,\nonumber
\eeq
where we neglected the imaginary part of $\varpi^{2\ell+2}$. 

In order to get the energy flux, we need to calculate now (\ref{transformationR}).  First of all, the solution (\ref{sol2F1G}) reads
\be
X_{\omega\ell m}(x)=A(1-x)^{\ell} x^{-i \varpi} \left[ F(\ell+3-i\varpi,\ell-1-i\varpi,1-2i\varpi,x)-\frac{x^{2i\varpi}}{x_0^{2 i\varpi}}F(\ell+3+i\varpi,\ell-1+i\varpi,1+2i\varpi,x)\right] \,,\label{exactsolution}
\ee
and from (\ref{nearregionG}) we find the asymptotic behaviour to be dominated by
\beq
%X(r_p)\sim A\left( \frac{r_H}{r_p}\right)^{\ell} \frac{\Gamma(\ell-1)}{2\Gamma(2\ell+2)} (-i\varpi) \Gamma(\ell+3)(1+x_0^{-i\varpi})
X(r_p) 
& = &  A\left[\frac{r_p}{r_H}\right]^{\ell+1}\Gamma(1+2\ell) \left(   \frac{1-x_0^{-2i\varpi}}{\Gamma(-1 + \ell) \Gamma(3 + \ell)} +\frac{i(2 \gamma_{EM} + \psi( -1 + \ell) + \psi( 3 + \ell)) (1+x_0^{-2i\varpi})  \varpi}{\Gamma(\ell-1)\Gamma(3+\ell)} \right) +...\nonumber
\eeq
where  "$...$" denotes higher order terms in $r_H/r_p$ and $\varpi$. Because $\varpi\approx \frac{n \pi}{|\log x_0|}$ the first term in the parenthesis is clearly subdominant, and we shall neglect it. In the BH case this would have been the dominant contribution, but because of the different boundary conditions now chosen, it is the subsequent leading term what dominates now.

On the other hand, we calculate the derivative from (\ref{exactsolution}), 
\beq
&&\frac{d}{d r}X|_{r_p}  =  \frac{2M}{r_p^2} \left[ -X\left(\frac{\ell}{1-x}+\frac{i\varpi}{x}\right)   -2i\varpi  A(1-x)^{\ell}\frac{x^{i\varpi-1}}{x_0^{2 i\varpi}}F(\ell+3+i\varpi,\ell-1+i\varpi,1+2i\varpi,x)  \right. \nonumber \\
&&\left. +A(1-x)^{\ell}x^{-i\varpi}   \left( \frac{(\ell+3-i\varpi)(\ell-1-i\varpi)}{1-2i\varpi} F(\ell+4-i\varpi,\ell-i\varpi,2-2i\varpi,x) - \frac{x^{2i\varpi}}{x_0^{2 i\varpi}} c. c.  \right) \right]\,,  \label{AuxApG}
  %\times  \left. \left(\frac{(\ell+3-i\varpi)(\ell-1-i\varpi)}{1-2i\varpi} F(\ell+4-i\varpi,\ell-i\varpi,2-2i\varpi,x) -c.c \right) \right] \nonumber
\eeq
and analyze the large distance behaviour using (\ref{hypergeometricproperties}). Taking care of the issues commented above, the leading order contribution is
\beq
\frac{d}{d r}X|_{r_p} &  = &    \frac{2M}{r_p^2}A \left[\frac{r_p}{r_H}\right]^{\ell+2} \frac{2i (2 \gamma_{EM} + \psi( -1 + \ell) + \psi( 3 + \ell)) \varpi}{\Gamma(\ell-1)\Gamma(3+\ell)}\Gamma(1+2\ell) (1+\ell) + ... 
\eeq
Consequently, from (\ref{transformationapprox}) we get
\beq
R^H_{\omega \ell m} & \approx & (\ell(\ell+1)+1) 2r_H A \left[\frac{r_p}{r_H}\right]^{l+2} \frac{2i(2 \gamma_{EM} + \psi( -1 + \ell) + \psi(3 + \ell)) \varpi}{\Gamma(\ell-1)\Gamma(3+\ell)}\Gamma(1+2\ell) \,.
\eeq   
Finally, from (\ref{Zimproved}), (\ref{b0}), (\ref{transformationB}) and \eqref{spherical_equator}, we get the energy flux after several simplifications:
\beq
|Z_{\ell m}|^2 & \approx & 
%  \frac{4\pi^2 \varpi^{2\ell+6} r_p^{-2\ell-2}  r_H^{2\ell-2} \Gamma(2\ell+2)\Gamma(\ell-1)^3\Gamma(\ell+3)^3\cos^2(\varpi \log x_0)}{ \Gamma(1+2\ell)^2 4^{2\ell+2}\Gamma(\ell+3/2)^2\Gamma(2\ell+2)^2 \sin^2(\varpi \log x_0) + \Gamma(-\ell+1/2)^2 \varpi^{4(\ell+1)} \Gamma(\ell+3)^4\Gamma(\ell-1)^4 \cos^2(\varpi \log x_0) }  \nonumber
\frac{4^2\pi^2 \varpi^{2\ell+6} r_p^{2\ell}  r_H^{-2\ell-4} \Gamma(2\ell+2)^3\Gamma(2\ell+1)^2 \Gamma(\ell-1)(\ell+1)^4 [2\gamma_{EM}+\psi(\ell-1)+\psi(\ell+3)]^2/\Gamma(\ell+1)^2/\Gamma(\ell+3)}{ \Gamma(1+2\ell)^2 4^{2\ell+2}\Gamma(\ell+3/2)^2\Gamma(2\ell+2)^2 \sin^2(\varpi \log x_0) + \Gamma(-\ell+1/2)^2 \varpi^{4(\ell+1)} \Gamma(\ell+3)^4\Gamma(\ell-1)^4 \cos^2(\varpi \log x_0) } \,.\nonumber\\ \label{fluxnearQNM2}
\eeq
Near the QNM frequencies, this is
\beq
\dot E_{s=2} & \approx & %\frac{2^{(2\ell-1)/3}m_0^2}{\pi^{(2\ell-1)/3} k^{2(\ell-2)/3} M^2} \frac{\Gamma(2\ell+2)\Gamma(\ell+1/2)^2}{\Gamma(\ell-1)\Gamma(\ell+3)}(\log\epsilon)^{2(\ell-2)/3}
  \frac{m_0^2}{M^2} \frac{m^{4\ell/3} }{(n \pi)^{10\ell/3}}(\log \epsilon)^{10\ell/3} \frac{2\pi      \Gamma(2\ell+2)^3\Gamma(2\ell+1)^2 (\ell+1)^4 [2\gamma_{EM}+\psi(\ell-1)+\psi(\ell+3)]^2}{2^{2\ell/3} \Gamma(-\ell+1/2)^2  \Gamma(\ell+3)^5\Gamma(\ell-1)^3 \Gamma(\ell+1)^2 }\,. \label{fluxpeak2}
\eeq
\end{widetext}

%%%%%%%%%%%%%%%%%%%%%%%%%%%%%%%%%%%%%%%%%%%%%%%%%%%%%%%%
\subsection{The resonance width}
%%%%%%%%%%%%%%%%%%%%%%%%%%%%%%%%%%%%%%%%%%%%%%%%%%%%%%%%
We follow the same method as in the scalar case. Assuming the frequency band to be narrow around the resonance modes, we can expand as in (\ref{bandw}). On the other hand,  the gravitational flux (\ref{fluxnearQNM2}) can be written as
\beq
\dot E_{s=2}(r) & = & \frac{ a_{\ell m}\varpi^{2\ell/3+4}(r)}{b_{\ell }\varpi^{4\ell+4}(r)\cos^2(\varpi(r)\log x_0)+c_{\ell } \sin^2(\varpi(r)\log x_0)} \,,\nonumber
\eeq
now with $b_{\ell}\equiv \Gamma(-\ell+1/2)^2\Gamma(\ell-1)^{4}\Gamma(\ell+3)^4>0$ and $c_{\ell}\equiv 4^{2\ell+2}\Gamma(2\ell+2)^2\Gamma(2\ell+1)^{2}\Gamma(\ell+3/2)^2>0$ ($a_{\ell m}$ will not be needed). The flux peaks at $r_{\rm QNM}$, and reads $\dot E_{s=2}(r_{\rm QNM})=\frac{a_{\ell m}}{b_{\ell}} \varpi^{-10/3\ell}_{\rm QNM}$. By expanding the energy flux up to second order we get
\begin{widetext}
\beq
\frac{\dot E_{s=2}(r)}{\dot E_{s=2}(r_{\rm QNM})}=\left[1+5\ell \frac{r-r_{\rm QNM}}{r_{\rm QNM}} +\left(\frac{r-r_{\rm QNM}}{r_{\rm QNM}} \right)^2\left(\frac{9n^2\pi^2-35\ell+50\ell^2}{4}-\frac{9c_{\ell} n^2 \pi^2 \varpi_0^{-4-4\ell}}{4b_{\ell}} \right) \right]+\dots
\eeq
\end{widetext}
Demanding now $\dot E_{s=2}(r) = \frac{1}{2} \dot E_{s=2}(r_{\rm QNM}) $ one gets a quadratic equation for $(r-r_{\rm QNM})/r_{\rm QNM}\equiv y$,
\beq
y^2\left[ \frac{9n^2\pi^2-35\ell+50\ell^2}{4}-\frac{9c_{\ell} n^2 \pi^2 \varpi_{\rm QNM}^{-4-4\ell}}{4b_{\ell}} \right]+y(5\ell)=-\frac{1}{2} \,,\nonumber
\eeq
which again leads to two solutions, $r_{+}$ and $r_{-}$,  corresponding to the outer and inner radius of the width. The solution, in the approximation in which $\varpi_{\rm QNM}\ll1$ (recall (\ref{QNMfreq})), takes the simple form:
\beq
y_{\pm} \approx \pm \frac{\varpi_{\rm QNM}^{2\ell+2}}{k\pi} \sqrt{  \frac{2 b_{\ell} }{9c_{\ell}}    }\,.
\eeq
Using (\ref{gravitationalF}) it can be  checked that $\sqrt{b_{\ell} /c_{\ell} }= F_{s=2}(\ell)(-1)^{\ell+1}$. Thus we finally arrive at
\beq
\left| \frac{r_+-r_-}{r_{\rm QNM}} \right| &\approx & \sqrt{\frac{8}{9} } \frac{ {\rm Im}\varpi_{\rm QNM}}{ {\rm Re} \varpi_{\rm QNM}}  \\
&\approx &  \sqrt{\frac{8}{9} }\frac{(2 n \pi)^{2\ell+1} \Gamma(\ell+1)^2\Gamma(\ell-1)^2\Gamma(\ell+3)^2 }{ {4} |\log x_0|^{2\ell+2} (2\ell+1) \Gamma(2\ell+1)^3}\,, \nonumber
\eeq
where we used the form of (\ref{widthaux}). The width in frequency follows the same ideas as in \eqref{freqwidthw}, yielding the same relation.

%%%%%%%%%%%%%%%%%%%%%%%%%%%%%%
\bibliographystyle{apsrev4}
\bibliography{References}

\end{document}